\documentclass[%
 reprint,
 amsmath,amssymb,
 aps,
]{revtex4-2}
\usepackage{graphicx}% Include figure files
\usepackage{dcolumn}% Align table columns on decimal point
\usepackage{bm}% bold math
\usepackage[table]{xcolor}

\begin{document}
\preprint{APS/123-QED}
\title{Tailoring Robust Quantum Anomalous Hall Effect via Entropy-Engineering}

\author{Syeda Amina Shabbir}
\affiliation{Institute for Superconducting and Electronic Materials (ISEM), Faculty of Engineering and Information Sciences (EIS), University of Wollongong, Wollongong, New South Wales 2525, Australia}
\author{Frank Fei Yun}
\affiliation{Institute for Superconducting and Electronic Materials (ISEM), Faculty of Engineering and Information Sciences (EIS), University of Wollongong, Wollongong, New South Wales 2525, Australia}
\author{Muhammad Nadeem}
\email{mnadeem@uow.edu.au}
\affiliation{Institute for Superconducting and Electronic Materials (ISEM), Faculty of Engineering and Information Sciences (EIS), University of Wollongong, Wollongong, New South Wales 2525, Australia}
\author{Xiaolin Wang}
\email{xiaolin@uow.edu.au}
\affiliation{Institute for Superconducting and Electronic Materials (ISEM), Faculty of Engineering and Information Sciences (EIS), University of Wollongong, Wollongong, New South Wales 2525, Australia}


\begin{abstract}
The development of quantum materials and the tailoring of their functional properties is of fundamental interest in materials science. Here, a new design concept is proposed for the robust quantum anomalous Hall effect via entropy engineering in 2D magnets. As a prototypical example, the configurational entropy of monolayer transition metal trihalide VCl$_3$ is manipulated by incorporating four different transition-metal cations [Ti,Cr,Fe,Co] into the honeycomb structure made of vanadium, such that all in-plane mirror symmetries, inversion and/or roto-inversion are broken. Monolayer VCl$_3$ is a ferromagnetic Dirac half-metal in which spin-polarized Dirac dispersion at valley momenta is accompanied by bulk states at the $\Gamma$-point and thus the spin-orbit interaction-driven quantum anomalous Hall phase does not exhibit fully gapped bulk band dispersion. Entropy-driven bandstructure renormalization, especially band flattening in combination with red- and blue-shifts at different momenta of the Brillouin zone and crystal-field effects, transforms Dirac half-metal to a Dirac spin-gapless semiconductor and leads to a robust quantum anomalous Hall phase with fully gapped bulk band dispersion and, thus, a purely topological edge state transport without mixing with dissipative bulk channels. These findings provide a paradigm for designing entropy-engineered 2D materials for the realization of robust quantum anomalous Hall effect and quantum device applications.
\end{abstract}

%\keywords{a,b}%Use showkeys class option if keyword
\maketitle

\section{Introduction}
Spin-gapless semiconductors (SGSs) \cite{wang08,nadeem24} are characterized by gapless dispersion in one of the spin sectors while a gapped spectrum in the other spin sector. With this unique characteristic, SGSs bridge the gap between magnetic semiconductors \cite{kossut93} and magnetic half-metals \cite{de83,van95}. SGSs also serve as a fundamental ingredient for theoretical understanding and experimental realization of various exotic quantum phases such as the quantum anomalous Hall (QAH) effect \cite{haldane88,nadeem20,nadeem24}, new (quantum) anomalous spin Hall effects \cite{wang17}, and topological nodal line spin-gapless semimetals \cite{ding22,nadeem24}. In addition, because of their intriguing electronic, magnetic and optical features, SGSs are promising candidates for quantum device applications, both with conventional bulk transport in tunneling junctions featuring large spin filtering and high tunneling magnetoresistance and with dissipationless topological edge state transport \cite{SGS-D19,SGS-D20,SGS-D22,Maji22,feng20,nadeem24}.

After the seminal proposal for SGSs in 2008 \cite{wang08}, SGSs have been predicted in a variety of materials classes \cite{nadeem24}. However, the experimental realization of SGSs and spin-orbit interaction (SOI) driven QAH effect has been hindered on several fronts. For instance, while indirect SGSs with parabolic dispersion have been confirmed, experimental confirmation of direct SGSs with Dirac/parabolic dispersion is limited. Similarly, other than intrinsic SGSs, various QAH materials have been fabricated and characterized, but with low Curie temperature, small topological bandgap, and the presence of dissipative channels along the dissipationless chiral edge states. These challenges can be evaded with experimentally accessible direct SGSs where intrinsic magnetism and SOI could open a topologically nontrivial bandgap featuring a robust QAH effect. The search for direct SGSs and the corresponding high-temperature (Curie) robust QAH effect motivates bandstructure engineering in half-metallic magnets and magnetic semiconductors.

\begin{figure*}[ht]
\centering
\includegraphics[width=0.8\linewidth]{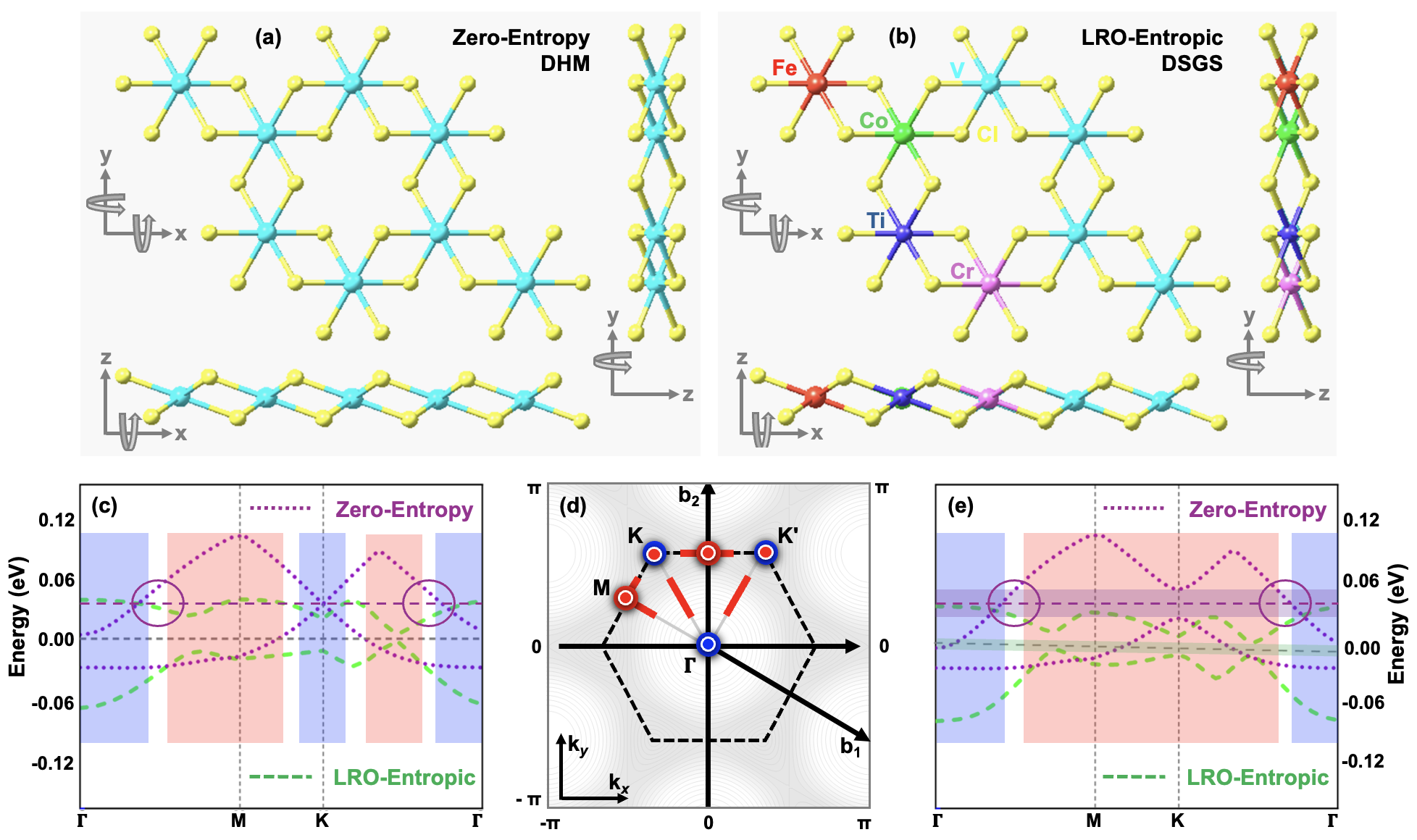}
\caption{\textbf{Design concept and bandstructure renormalization via entropy engineering.} \textbf{(a, b)} Lattice structure of zero-entropy V$_8$Cl$_{24}$ monolayer (a) and entropic V$_4$TiCrFeCoCl$_{24}$ monolayer (b) with top and side views. \textbf{(c-e)} Low-energy Dirac bands of zero-entropy V$_8$Cl$_{24}$ monolayer (purple) and entropic V$_4$TiCrFeCoCl$_{24}$ monolayer (green) without SOI (c) and with SOI (e). Here, red and blue regions, respectively, represent entropy-driven red and blue shifts in the energy spectrum. The first Brillouin zone showing red and blue shifts at/along high-symmetry points/lines where large (small) spheres indicate red/blue shift in the absence (presence) of SOI (d). Although the energy shift remains red and blue across the M and $\Gamma$ points, respectively, the energy dispersion across valleys K/K$^\prime$ exhibits a blue shift in the absence of SOI while a red shift in the presence of SOI. These entropy-driven red and blue shifts in the energy spectrum lead to the bandstructure renormalization.}
\label{HESGS}
\end{figure*}

Here we propose a new design concept for SGSs through entropy engineering. Entropy engineering is a new concept of materials design rendering the entropy-dominated phase stabilization and tailoring of functional properties \cite{murty19-book,sarkar19-AM,hsu24-NRC,schweidler24-NRM}. Due to the direct bandgap closing and an inherent connection of Dirac half-metals (DHMs) with Dirac SGSs (DSGSs), we consider monolayer vanadium trichloride VCl$_3$ as a material platform and manipulated its configurational entropy by incorporating four different transition metal (TM) atoms [M$^\prime$: Ti, Cr, Fe, Co] in the honeycomb structure formed by vanadium cations, as shown in figure \ref{HESGS}(a,b). The M$^\prime$ cations are substituted by vanadium atoms such that all in-plane mirror symmetries, inversion, and/or roto-inversion, are broken. Entropy-dominated effects on the electronic and spintronic properties of entropic V$_{0.5}$(TiCrFeCo)$_{0.5}$Cl$_3$ monolayer are studied using first-principles calculations. The entropy manipulation results in several intriguing features, such as band flattening due to red and blue shifts at different momenta of the Brillouin zone and a momentum shift of Dirac points due to the modification of in-plane crystal fields, as shown in Figure \ref{HESGS} (c, d, e). As a consequence of band flattening and crystal-field effects, VCl$_3$ monolayer undergoes a major bandstructure renormalization. First, renormalization of the entropy-driven band structure transforms the DHM phase in the VCl$_3$ monolayer to a DSGS phase in the entropic V$_{0.5}$(TiCrFeCo)$_{0.5}$Cl$_3$ monolayer, referred here as entropy-engineered DSGS (EE-DSGS). Second, when SOI is activated, a nontrivial bandgap leads to a robust QAH phase in V$_{0.5}$(TiCrFeCo)$_{0.5}$Cl$_3$ monolayer. The robustness of the QAH phase is derived from the bandstructure renormalization; a maximal blue shift at the $\Gamma$-point disentangles chiral edge states lying in the QAH gap from the bulk states and thus assures dissipationless topological transport via chiral edge states.

\section{Methods and Results}
All the computations are performed using the Vienna ab initio simulation package (VASP) \cite{kresse93,kresse99} within the generalized gradient approximation (GGA), employing the Perdew-Burke-Ernzerhof (PBE) exchange correlation functional \cite{perdew96}. Electronic and nucleic interactions are explained using the projector augmented wave (PAW) \cite{blochl94} technique. The energy criterion is set to $10^{-5}$ ev, while the atom force convergence is equal to 0.01 eV \AA$^{-1}$. Moreover, a PW (plane wave) kinetic energy cutoff is set to 500 eV. A 2 x 2 x 1 supercell is built with vacuum spacing equal to 15 \AA along the normal direction so that inter-layer interaction is prevented. The Brillouin zone was sampled using a 2 x 2 x 1 and 11 × 11 x 1 Gamma-centered Monkhorst–Pack grids \cite{monkhorst76} for optimization and electronic structure calculations. The unfolding of bands in VCl$_3$ 2 x 2 supercell is performed using the VASP band unfolding package, VaspBandUnfolding \cite{QZheng}. In addition, orbital-resolved band structures are calculated using vaspkit software \cite{wang21} and wanniertools software package is used to calculate edge states \cite{wu18}.

\begin{figure*}[ht]
\centering
\includegraphics[width=0.8\linewidth]{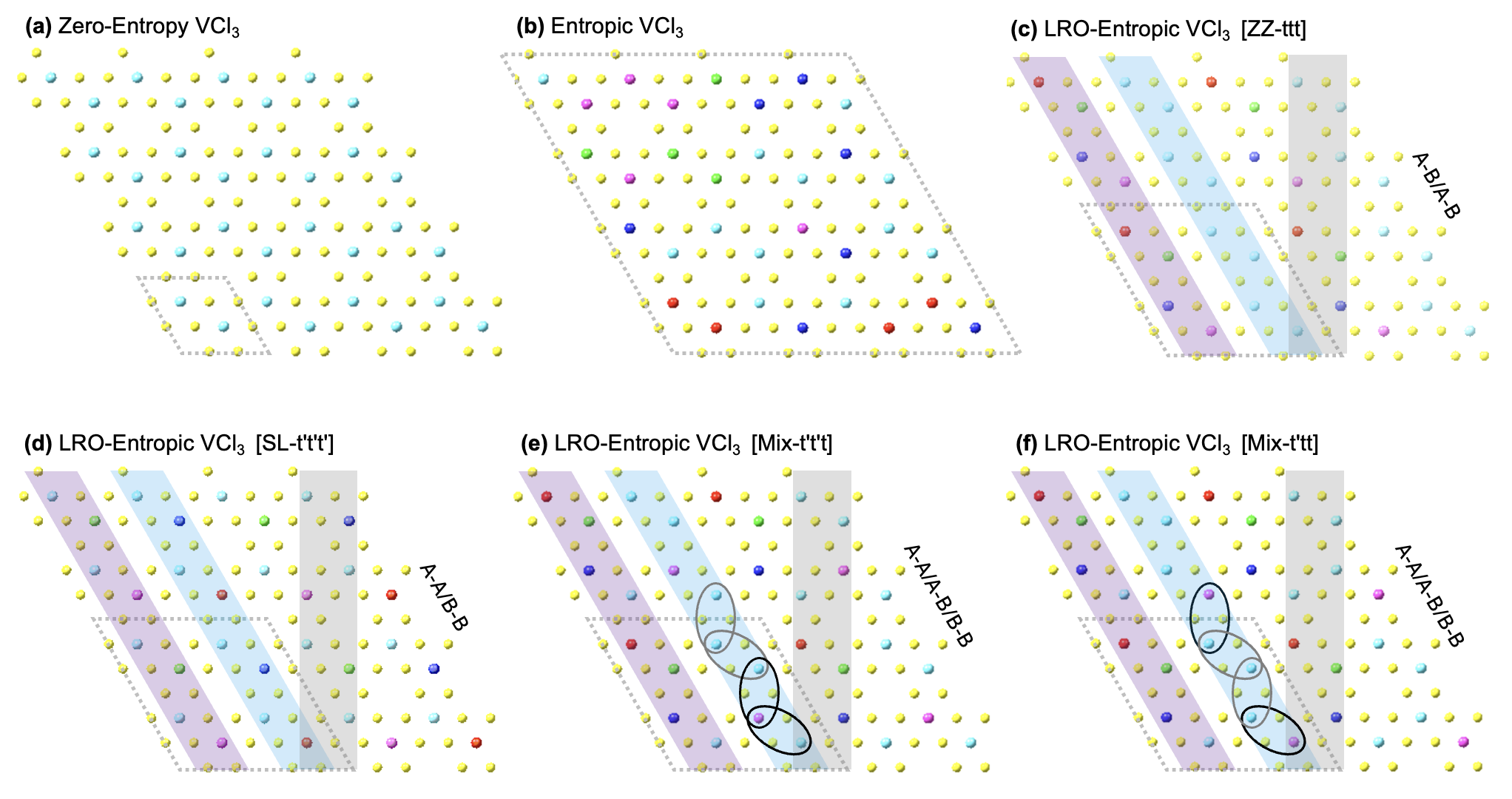}
\caption{\textbf{Entropic VCl$_3$ monolayer with a long-range order.} \textbf{(a,b)} A 4 x 4 supercell of zero-entropy VCl$_3$ (a) and maximal entropic VCl$_3$ with a random distribution of TM atoms at A and B sublattice sites of honeycomb structure, without a long-range order (b). \textbf{(c,d)}. A 4 x 4 supercell of LRO-entropic VCl$_3$ with a long-range order along zigzag chains [ZZ-ttt] with A-B/A-B bonds favored by V/M$^\prime$ atoms (c) and a long-range order on sublattice sites [SL-t$^\prime$t$^\prime$t$^\prime$] with A-A/B-B bonds favored by V/M$^\prime$ atoms (d). \textbf{(e,f)} A 4 x 4 supercell of LRO-entropic VCl$_3$ with a mixed preference of A-A/A-B/B-B bonds by V/M$^\prime$ atoms, [Mix-t$^\prime$t$^\prime$t] (e) and [Mix-t$^\prime$tt] (f). Panel (f) is same as panel (e), with an exchange of V and Cr sublattice sites, enclosed by black ovals. Here, t and t$^\prime$, respectively, represent nearest-neighbor hopping amplitude along V-V/M$^\prime$-M$^\prime$ bonds and the V-M$^\prime$ bonds. The t-triplet represents the pattern of nearest-neighbor hopping along zigzag chains, represented by blue and purple strips. Here, vertical gray strips represent the armchair chains.}
\label{LRO}
\end{figure*}

Configurational entropy ($S_{\text{conf}}$) is a controllable parameter, and the `level of entropy' within the high entropy matrix could be tuned to achieve desired system functionalities. That is, an interplay between a global disorder and a local disorder allows to explore materials' space exhibiting on-demand functional properties. It is important to note that neither a high configurational entropy measuring a degree of disorder nor a compositional complexity relying on diverse cations/anions could necessarily enhance functionalities, but a local environment determines material's electronic and magnetic properties via local interactions. In a high entropy configuration, the cation sites are randomly occupied by five or more TM atoms with equiatomic ratios such that $S_{\text{conf}}\ge1.5R$, where $R$ is the gas constant. The presence of clustering, phase separation, short-range ordering (SRO), or long-range ordering (LRO), lowers the entropy level from its maximal value, which is due to complete randomness, to its intermediate ($1R\ge S_{\text{conf}}<1.5R$) or low ($0<S_{\text{conf}}<1R$) values.

The `level of entropy' and the type of ordering may depend upon the interplay between kinetic (favorable activation energy) and thermodynamic (favorable Gibbs free energy) control of the phase transition from the synthesis temperature to the thermodynamic equilibrium at room temperature \cite{aamlid23}. A high entropy system typically synthesized at a high temperature could be stabilized to yield the desired final state the system exhibits by optimizing the thermal history; synthesis temperature, cooling rate, and equilibration time. That is, high entropy of a multi-component system with random atomic occupancy could be retained through quenching. However, depending on the interatomic interactions, the intermediate equilibration time allows the formation of clustering or SRO, while a longer equilibration time could potentially cause LRO or complete phase separation. In either case, the emergence of the quantum phases inherently dependent on the nearest-neighbor hopping or local interaction could potentially be realized through bandstructure renormalization via entropy engineering in a multi-component system. As a seminal case study on the role of entropy engineering in topological quantum matter, here we present a proof-of-concept to demonstrate bandstructure renormalization and the emergence of novel quantum phases in multi-component 2D materials with long-range ordering.

We consider the VCl$_3$ monolayer as a prototypical 2D half-metallic ferromagnet and its configurational entropy is increased by substituting four vanadium atoms with TM atoms [M$^\prime$: Ti, Cr, Fe, Co], as shown in figure \ref{LRO}. The highest level of entropy is traditionally derived from a random distribution of TM atoms at the A and B sublattice sites of the honeycomb structure, figure \ref{LRO}(b). However, the entropy level of V$_{0.5}$(TiCrFeCo)$_{0.5}$Cl$_3$ monolayer lowers from its maximal value when the system stabilizes with the long-range ordering. In this article, as shown in figure \ref{LRO}(c-f), four distinct long-ranged ordered entropic configurations of V$_{0.5}$(TiCrFeCo)$_{0.5}$Cl$_3$ monolayer are studied, displaying an ordering of V/M atoms along (i) zigzag chains with A-B/A-B bonds favored by V/M$^\prime$ atoms, (ii) sublattice sites with A-A/B-B bonds favored by V/M$^\prime$ atoms, and (iii-iv) a long-range ordered entropic VCl$_3$ with a mixed preference of A-A/A-B/B-B bonds by V/M$^\prime$ atoms. These LRO-entropic configurations are labeled as ZZ-ttt, SL-t$^\prime$t$^\prime$t$^\prime$, Mix-t$^\prime$t$^\prime$t, and Mix-t$^\prime$tt where t and t$^\prime$ represent the amplitude of nearest-neighbor hopping along V-V/M$^\prime$-M$^\prime$ bonds and V-M$^\prime$ bonds, respectively. The t-triplet represents a pattern of nearest-neighbor hopping between sublattice sites along zigzag chains.

\subsection{VCl$_3$ Monolayer}
The monolayer VCl$_3$ possess a buckled trilayered structure Cl-V-Cl with V atoms sandwiched between Cl atoms, where V atoms form a hexagonal honeycomb structure in which V$^{3+}$ cations are covalently bonded to six nearest neighboring Cl$^-$ anions. Each of the Cl$^-$ anions is bonded to two V$^{3+}$ cations via edge sharing octahedral coordination. Full geometry optimization is carried out on the unit cell of the unchanged VCl$_3$ monolayer to relax the structure. The optimized lattice constant obtained is 6.28 \AA, which is in good agreement with those previously reported for VCl$_3$ monolayer \cite{he16,zhou16,feng20,ouettar23} and VCl$_3$ crystal \cite{klemm47}.

Several theoretical studies, along with recent experimental observations \cite{mastrippolito23,deng25}, have been reported on VCl$_3$. However, mainly due to potentially different and competing ground states associated with different underlying crystalline symmetries and external stimuli \cite{camerano25,camerano24}, the VCl$_3$ monolayer could display a different electronic structure and magnetic ordering. First-principles calculations show that the VCl$_3$ monolayer could be a ferromagnetic half-metal \cite{he16,zhou16,feng20, ouettar23}, a ferromagnetic insulator \cite{tomar19}, and a Mott-Hubbard insulator \cite{mastrippolito23} with a ferromagnetic Mott insulating ground state followed by an antiferromagnetic transition at $T_N=21.8$ K. The seminal experimental study \cite{mastrippolito23} that reported the synthesis and characterization of single-crystalline VCl$_3$, accompanied by experimental observations and first-principles calculations to understand the electronic nature, suggested an intrinsic Mott-Hubbard insulating phase and an extrinsic 2D magnetic polaronic phase. It shows that even a small perturbation could transform an extremely correlated Mott insulating phase into a 2D polaronic phase in the VCl$_3$ ionic system through a band inversion between spin-polarized $a_{1g}$ and $e^\prime_g$ V-3d states. The band inversion is determined by an upward energy shift of the $e^\prime_g$ states in the valence bands and the crossing of the $a_{1g}$ states from the conduction to the valence bands, so that both $a_{1g}$ and $e^\prime_g$ states fall into the valence bands with an inverted energy position.

Interestingly, unlike the phase transition driven by internal correlation and external stimuli in other monolayers of transition metal trihalides, half-metallic and insulating phases in VCl$_3$ could be classified by Fermi level manipulation. For example, in OsCl$_3$ exhibiting the SGS phase \cite{OsCl}, correlation-driven topological phase transition from the QAH to the Mott insulating phase is implemented by closing and reopening the bandgap in the manifold of Os t$_{2g}$ bands crossing the Fermi energy. However, the distinction between a half-metallic phase and an insulating phase in VCl$_3$ is more related to the Fermi level position than the closing and reopening of the bandgap. That is, the low-energy manifold of V-d bands across the Fermi energy is characterized by a spin-polarized Dirac point in both the half-metallic and insulating phases. When the dispersing Dirac bands cross the Fermi level, the system is classified as a half-metal \cite{he16,zhou16,feng20,ouettar23}. However, when the Dirac bands lie below the Fermi level, with a valence band maximum (VBM) touching the Fermi level at the M-point, the system is classified as an insulator \cite{tomar19,mastrippolito23}. More recently, using first-principles methods, L. Camerano et al \cite{camerano25,camerano24} showed that the monolayer VCl$_3$ realizes a ground state with multi-component magneto-orbital order. The multi-component ground state with simultaneous magnetic and orbital ordering explicitly demonstrates the underlying symmetry-determined origin of the half-metallic and insulating phases in monolayer VCl$_3$. The distribution of low-energy d orbitals, the symmetry determination of the ground state, and the entropy-driven effects on the electronic phases of VCl$_3$ are further elaborated in Section 2.3 which explicitly demonstrates the orbital-resolved band dispersion.

Based on our first-principles calculations, consistent with the underlying symmetry subject to the Fermi-level pinned at zero-energy level ($E_F=0$), the pristine VCl$_3$ monolayer is a ferromagnetic DHM \cite{he16,zhou16}; characterized by a Dirac dispersion in the spin-up channel and a large indirect bandgap of $>4.5$ eV in the spin-down channel, and an intrinsic long-range ferromagnetic character with estimated Curie temperatures up to 425 K \cite{zhou16}. Due to its DHM character, VCl$_3$ is predicted to exhibit the QAH effect \cite{he16,li18,liu18}. However, despite all these interesting electronic and magnetic properties that are supposed to provide an ideal playground for the QAH effect, the VCl$_3$ monolayer does not guarantee a robust QAH phase, which is evident from the spin-polarized band dispersion, as shown in figure \ref{HESGS}(c,e). The Dirac point lying above the Fermi level at valley momenta (K/K$^\prime$) is accompanied by bulk states crossing the Fermi level at the $\Gamma$-point. This behavior has also been confirmed through the non-vanishing density of states at the Fermi level \cite{zhou16}. As a result, as shown in the figure \ref{HESGS}(e), SOI opens a nontrivial bandgap at valleys but the bulk states at the $\Gamma$-point are further lowered and thus mix with the chiral edge states lying inside the nontrivial QAH gap. That is, even though spin-orbit coupled VCl$_3$ monolayer exhibits the QAH phase with a SOI-induced bandgap of 29 meV in the vicinity of valleys \cite{he16}, the QAH phase in VCl$_3$ monolayer does not remain robust due to mixing of edge states with dissipative bulk states, a problem that also remains inevitable in magnetic doped topological insulators \cite{nadeem20,Babar19,nadeem24}.

We also performed first-principles calculations for a 2 x 2 supercell of VCl$_3$ monolayer that is comprised of four unit cells, as shown in figure \ref{HESGS} (a). The electronic and magnetic properties of VCl$_3$ 2 x 2 supercell remain same as that obtained for VCl$_3$ primitive cell, as shown in figure \textbf{S1}. That is, spin-polarized band dispersion of VCl$_3$ 2 x 2 supercell also exhibits a ferromagnetic Dirac half-metallic character, though with several additional bands. In addition, consistent with calculations performed over VCl$_3$ primitive cell, SOI opens a nontrivial QAH bandgap at valley momenta, which is the same in magnitude to that obtained for VCl$_3$ primitive cell. However, similar to the VCl$_3$ primitive cell, bulk states in the vicinity of $\Gamma$-point lie inside the QAH gap. It reveals that the band dispersion of VCl$_3$ 2 x 2 supercell is mere a folding of VCl$_3$ 1 x 1 unit cell. This was further confirmed by unfolding the 2 x 2 band structure to reproduce the original VCl$_3$ band structure by using a VASP band unfolding procedure. The consistency of folding and unfolding of band dispersion in momentum-space is consistent with extrapolation in the real-space, i.e., 2 x 2 supercell of VCl$_3$ is simply an extrapolation of VCl$_3$ unit cell along x- and y-directions, as shown in figure \ref{HESGS} (a).

\begin{figure*}[ht]
\centering
\includegraphics[width=0.8\linewidth]{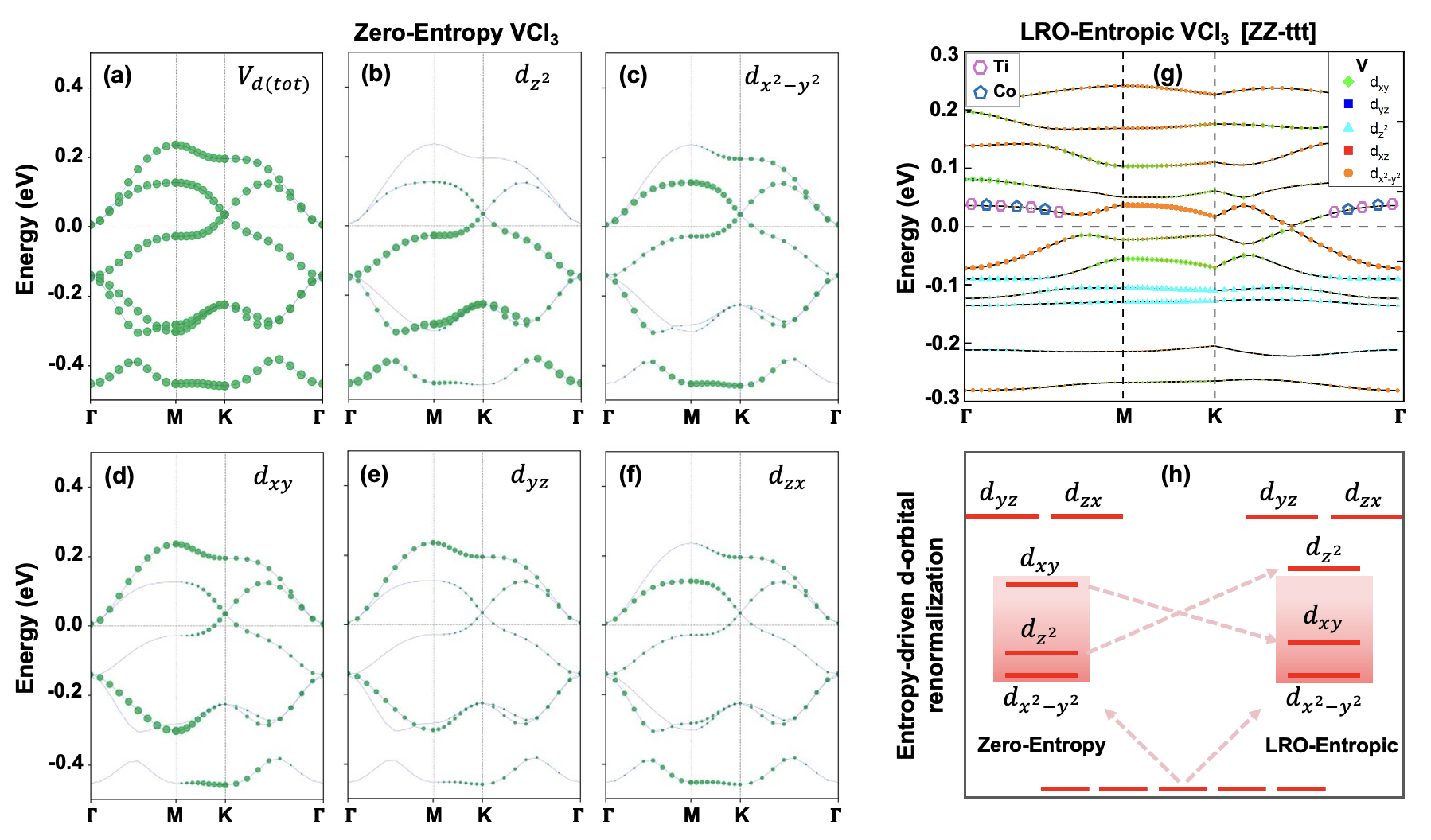}
\caption{\textbf{2D orbital-resolved electronic dispersion and bandstructure renormalization.} \textbf{(a-f)} Orbital-resolved band dispersion of pristine VCl$_3$ monolayer (1 x 1 unit cell) showing total contributions of 3d-orbitals of vanadium (a), contribution from $e_g$ (d$_{z^2}$, d$_{x^2-y^2}$) orbitals of vanadium (b,c), and the contribution from $t_{2g}$ (d$_{xy}$, d$_{yz}$, d$_{zx}$) orbitals of vanadium (d,e,f). \textbf{(g)} Orbital-resolved band dispersion of LRO-entropic VCl$_3$ monolayer in ZZ-ttt configuration (2 x 2 supercell) showing the contribution of 3d-orbitals of vanadium (V) and M$^\prime$ cations (Ti and Co). \textbf{(h)} A schematic representation of entropy-driven band-inversion between d$_{xy}$ and d$_{z^2}$ leading to the renormalization of vanadium 3d-orbital. The size of the bands represents the orbital weight.}
\label{ZZ}
\end{figure*}

\subsection{Entropic V$_{0.5}$(TiCrFeCo)$_{0.5}$Cl$_3$ Monolayer}
In the LRO-entropic VCl$_3$ with LRO along zigzag chains [ZZ-ttt], as shown in figure \ref{HESGS} (b) and \ref{LRO} (c), the M$^\prime$ atoms are incorporated along zigzag chains such that the lattice structure is constituted by adjacent zigzag chains made of V-atoms and M$^\prime$-atoms (Ti, Cr, Fe, Co) respectively. The geometric optimization of entropic V$_{0.5}$(TiCrFeCo)$_{0.5}$Cl$_3$ 2 x 2 supercell shows that the optimized lattice constants remain the same as that of pristine VCl$_3$ 2 x 2 supercell. However, the atomic positions/coordinates of both TM and Cl atoms are considerably changed. As shown in figure \ref{HESGS} and \ref{ZZ}, the entropy-driven bandstructure renormalization is featured by several interesting aspects such as band flattening, red and blue shifts at different momenta of the Brillouin zone, and a momentum shift of Dirac points. In the absence of SOI, figure \ref{HESGS}(c) and \ref{ZZ}(g), band flattening is a consequence of a blue shift in the vicinity of high-symmetry $\Gamma$-point and valleys K/K$^\prime$ while a red shift around the M-point and momenta along the K$-\Gamma$ symmetry line. While the blue shift opens a trivial gap at the Dirac points K/K$^\prime$, the red shift at momenta k and k$^\prime$ along the symmetry lines K$-\Gamma$ and K$^\prime-\Gamma$ induces Dirac points at the Fermi level. As a result, Dirac points move away from valleys (K/K$^\prime$) indicating an entropy-driven renormalization of crystal field effects.

In addition, unlike the zero-entropy VCl$_3$ monolayer in which the Dirac point lies above the Fermi level at the K-point while additional bulk states cross the Fermi level at the $\Gamma$-point, the Dirac point in the LRO-entropic VCl$_3$ monolayer [ZZ-ttt] lies exactly at the Fermi level (with a negligible gap of 6.6 meV) and the low-energy states at the $\Gamma$-point move away from the Fermi level. As a result, with an increase in entropy, VCl$_3$ monolayer is transformed from a DHM to a DSGS. The remarkable aspect of doping is that, while the entropy of the material has increased, the fermi energy has decreased, sustaining the SGS Dirac point. This indicates that the electron density would have decreased. This would likely be a result of the electron localization around the doped ions. This is confirmed through the calculated Fermi energy and the bader charges.

\begin{figure}[ht]
\centering
\includegraphics[width=1\linewidth]{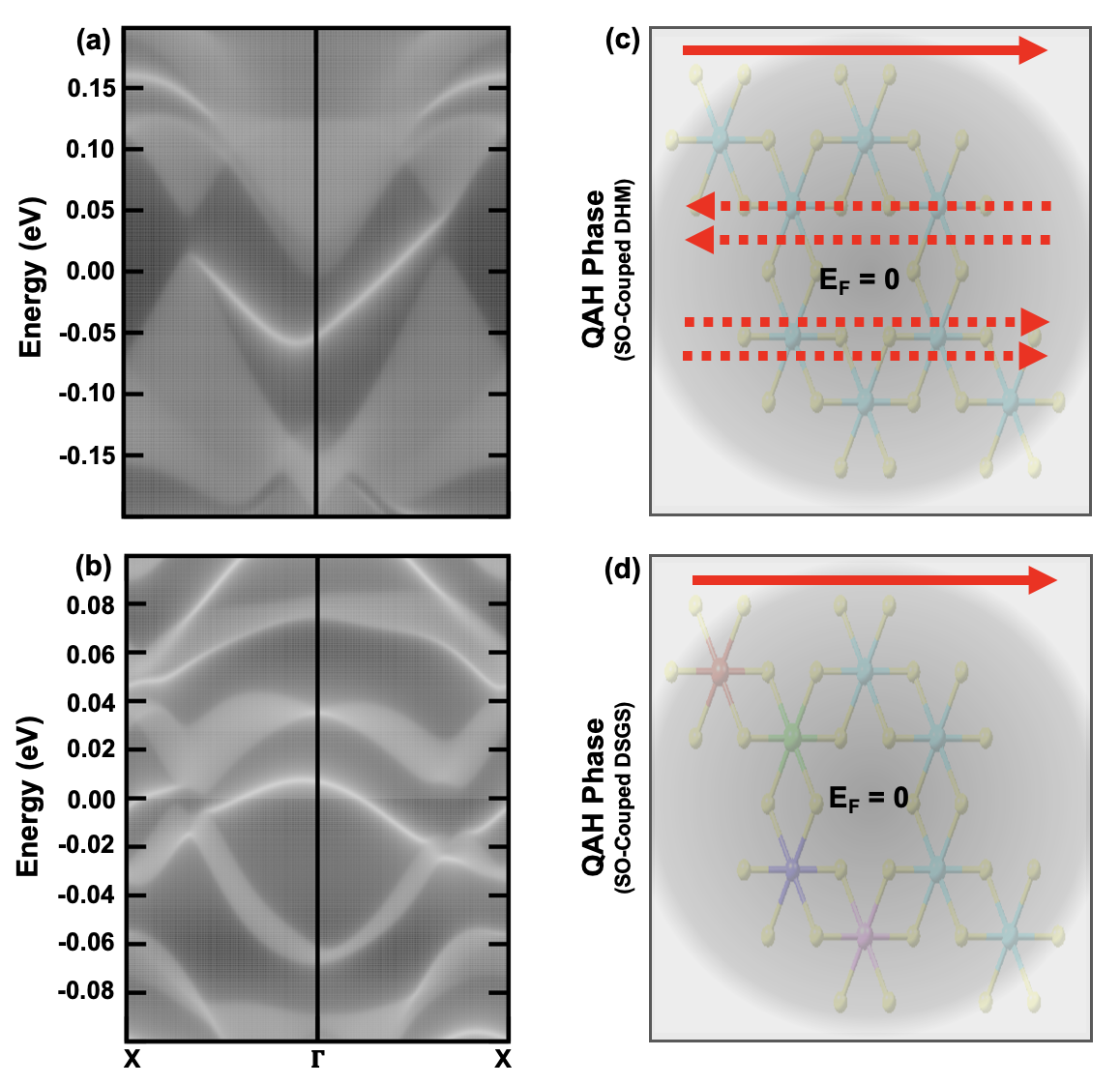}
\caption{\textbf{1D electronic dispersion and bulk-boundary correspondence characterizing QAH effect.} \textbf{(a, b)} 1D band structure of zero-entropy VCl$_3$ monolayer (1x1 unit cell) (a) and LRO-entropic VCl$_3$ [ZZ-ttt] monolayer (2x2 supercell) (b) showing the chiral edge state in momentum space along $\overline{X}-X-\Gamma$. \textbf{(c, d)} A real-space schematic representation of the chiral edge state (solid arrow) and bulk modes (dashed arrow) crossing the Fermi level $E_f=0$ in zero-entropy VCl$_3$ monolayer (c) and LRO-entropic VCl$_3$ [ZZ-ttt] monolayer (d). 1D dispersion clearly shows that the chiral edge states are mixed with the bulk modes in the zero-entropy VCl$_3$ monolayer (a,c), while the chiral edge states are completely disentangled from the bulk modes in the LRO-entropic VCl$_3$ [ZZ-ttt] monolayer (b,d), a key feature of entropy-engineering leading to a robust QAH effect.}
\label{1DBS}
\end{figure}

SOI opens a nontrivial bandgap at the k/k$^\prime$ points along the K/K$^\prime-\Gamma$ symmetry line featuring the QAH effect, as shown in figure \ref{HESGS}(e) and \textbf{S2}. The nontrivial topological character of SOI driven bandgap could be supported by various established mechanisms. First, the emergence of chiral edge states in 1D dispersion, figure \ref{1DBS}(b), validates the nontrivial bulk-boundary correspondence characterizing bulk states by non-vanishing Chern number $\mathcal{C}=\pm1$. Second, the origin of QAH is due to the intrinsic topology of the honeycomb lattice structure terminating on the zigzag edges \cite{nadeem22}. In the absence of SOI, the intrinsic topology of pristine honeycomb leads to zero-energy edge states connecting the two valleys K and K$^\prime$, while the bulk spectrum remains gapless. A finite SOI, acting as a perturbation, opens a gap in the 2D bulk spectrum and disperses the zero-energy modes into chiral edge states in the 1D edge-state spectrum. As shown in figure \ref{1DBS}, the chiral edge states connecting the two valleys confirm the intrinsic origin of the QAH effect in honeycomb lattice structures terminating on the zigzag edges. Third, the underlying physics of the QAH effect in spin-orbit coupled DSGSs is the same as that of the spinfull Haldane model \cite{nadeem24,nadeem20}, where SOI acts as a Haldane-type next-nearest-neighbor tunneling ($\lambda\tau\sigma_z$) \cite{haldane88}, a single copy of KM-type SOI ($\lambda\tau\sigma_zs_z$) in honeycomb lattice structures. As a result, the SOI-driven bandgap in the DSGSs leads to the spin QAH phase. Fourth, since the spin-gapless dispersion of the LRO-entropic VCl$_3$ monolayer [ZZ-ttt] is mainly constituted by vanadium d-orbitals around the Dirac point(s), the origin of the QAH effect is also consistent with the previously reported QAH effect in the zero-entropy VCl$_3$ monolayer \cite{he16,li18,liu18}.

Interestingly, unlike the half-metallic VCl$_3$ monolayer in which the spin-polarized Dirac dispersion at the valley momenta is accompanied by bulk states at the $\Gamma$-point and thus the QAH phase does not remain fully gapped, figure \ref{1DBS}(a), the LRO-entropic VCl$_3$ monolayer [ZZ-ttt] allows a fully gapped QAH phase and thus a purely topological edge state transport without mixing with dissipative bulk channels. The realization of DSGS phase and the robustness of corresponding QAH phase are associated with an entropy-driven band flattening in combination with red and blue shifts at different momenta of the Brillouin zone. However, unlike a blue shift in the absence of SOI, an increase in entropy induces a red shift around valley momenta in the spin-orbit coupled case. In addition, SOI has to overcome a gap of 6.6 meV in the pristine case before entering the QAH phase, more like a phase transition from magnetic semiconductor to the QAH phase. As a result, the size of the nontrivial bandgap in the entropic case is smaller than the bandgap in the zero-entropy case. In a fully gapped QAH phase in LRO-entropic VCl$_3$, the bulk bandgap could be tuned via a suitable selection of TM atoms with the assurance of a blue shift at both $\Gamma$ and K/K$^\prime$ points.

To further investigate entropy-driven bandstructure renormalization and the effects of symmetry breaking, LRO-entropic cases SL-t$^\prime$t$^\prime$t$^\prime$, Mix-t$^\prime$t$^\prime$t, and Mix-t$^\prime$tt are also analyzed. As shown in figure \ref{LRO}(d-f), along with the redistribution of TM atoms, nearest-neighbor hopping matrix elements are also modified. As a consequence, the electronic dispersion also gets drastically modified from that of the LRO-entropic ZZ-ttt case. As shown in figure \textbf{S2}, the LRO-entropic chiral structure SL-t$^\prime$t$^\prime$t$^\prime$ displays a gapped dispersion also in the spin-up sector and thus featuring a ferromagnetic insulating phase, Mix-t$^\prime$t$^\prime$t displays a nodal-line semi-metallic character between nearly flat spin-up and spin-down low-energy bands, while Mix-t$^\prime$tt is a ferromagnetic metal with a nearly flat spin-up band across Fermi level. The Mix-t$^\prime$t$^\prime$t configuration, with a nodal-line semi-metallic character, could also exhibit the QAH phase due to a Rashba-type SOI \cite{nadeem24}. A finite Rashba SOI could potentially be allowed through a renormalization of crystal field effects due to a modified atomic arrangement of different TM-atoms. Further details on the electronic and magnetic properties of these LRO-entropic cases are discussed in the Supplementary Information.

\subsection{Orbital-Resolved Band Dispersion}
In order to understand the microscopic origin of entropy-driven bandstructure renormalization, especially band flattening leading to the transition from half-metallic character to spin gapless semiconducting behavior and the accompanied effects on ferromagnetic ground state, orbital resolved band dispersion of LRO-entropic V$_{0.5}$(TiCrFeCo)$_{0.5}$Cl$_3$ monolayer [ZZ-ttt] is investigated, as shown in figure \ref{ZZ}. Due to the entropy-induced symmetry breaking, the bandstructure renormalization could originate from various effects at the microscopic level such as energy-splitting and redistribution of V-3d orbitals, contribution of spin-up electrons from the substituted TM-dopants, p-d hybridization between orbitals of Cl-atoms and TM-atoms, and the contribution of various TM-atoms to the total moment of magnetic ground state.

The pristine structure of VCl$_3$ stabilizes in ferromagnetic ground state with a total magnetic moment of $\approx 4.0$ $\mu_B$ per unit cell, where each V atom contributes $\approx 2.0$ $\mu_B$ while the magnetic moment of the surrounding Cl atoms is negligible. The ferromagnetic nature remains persistent for the 2 x 2 supercell displaying a total magnetic moment of $\approx 16.0$ $\mu_B$. These electronic and magnetic properties of the pristine monolayer VCl$_3$ are consistent with previously reported first-principle calculations \cite{he16,zhou16,ouettar23}. Like the half-metallic VCl$_3$ monolayer, the ground state of the LRO-entropic monolayer [ZZ-ttt] remains ferromagnetic; however, the total magnetic moment is reduced to 10.9729 $\mu_B$. The reduction in total magnetic moment is consistent with a phase transition from the DHM phase to a DSGS phase, in an itinerant ferromagnetic ground state.

As shown in figure \ref{ZZ}(a-f) and \textbf{S3}, the intrinsic half-metallicity in the VCl$_3$ monolayer is dominated by the vanadium 3d orbitals, with a negligible contribution from the Cl-p orbitals. The low-energy Dirac bands are predominantly occupied by $e_g$ orbitals ($d_{z^2}$ and $d_{x^2-y^2}$) and a $d_{xy}$ orbital with a contribution from the $d_{yz}$ and $d_{zx}$ orbitals around the $\Gamma$-point. The orbital contribution to low-energy Dirac bands is in good agreement with previous first-principles studies on the VCl$_3$ monolayer \cite{he16,ouettar23}. In addition to octahedral crystal field splitting of 3d-orbitals into a doublet $e_g$ ($d_{z^2}$ and $d_{x^2-y^2}$) and a triplet $t_{2g}$ ($d_{xy}$, $d_{yz}$ and $d_{zx}$), transition-metal trihalide monolayers with P31m symmetry could exhibit further splitting in the $t_{2g}$ triplet such that low energy states are contributed by a singlet $e_{out}\rightarrow a_{1g}\sim d_{z^2}$ and a doublet $d_{in}\rightarrow e^\prime_{g}\sim(d_{xy},d_{x^2-y^2})$ \cite{camerano25,camerano24,yang20}.

Like zero-entropy VCl$_3$ monolayer, spin-up bands of LRO-entropic monolayer [ZZ-ttt] are also predominantly occupied by a doublet $d_{in}\rightarrow e^\prime_{g}$ and a singlet $e_{out}\rightarrow a_{1g}$, as shown in figure \ref{ZZ} (g). However, there are several interesting modifications induced by enhanced configurational entropy. First, the low-energy Dirac bands are constituted by doublet $d_{in}\rightarrow e^\prime_{g}$ while the contribution of $e_{out}\rightarrow a_{1g}$ to the Dirac bands is completely depleted. It shows that only in-plane 3d orbitals ($d_{in}\rightarrow e^\prime_{g}$) contribute to the spin-up Dirac bands around the Fermi level, while the contribution from 3d-orbitals that extend along the out-of-plane direction ($d_{yz}$, $d_{zx}$, $d_{z^2}$) is diminished with an increase in the entropy, leading to a red shift around the M-point and along the K-$\Gamma$ line. The entropy-driven redistribution of 3d-orbitals is depicted in figure \ref{ZZ} (h). Second, the contribution from V atoms to a low-energy conduction band is significantly reduced at the $\Gamma$-point. On the other hand, as displayed in figure \ref{ZZ}(g) and \textbf{S4}(b, c), contribution of Ti and Co to the low-energy conduction band is prominent at the $\Gamma$-point. However, as shown in figure \textbf{S4}(d,e), the contribution from Fe and Cr atoms is vanishingly small to the low-energy valence bands. As a result, the bulk states move away from the Fermi level; a blue shift at the $\Gamma$-point, leading to the DHM-DSGS phase transition. Such an entropy-driven depletion of V-3d orbitals with weight along the z-axis and a finite contribution from doped M$^\prime$ atoms (Ti and Co) causes a band flattening due to red and blue shifts at various momenta of the Brillouin zone. 

The entropy-driven redistribution of 3d-orbitals and the accompanied red and blue shifts leading to the band flattening, is consistent with a symmetry-identified ground state in VCl$_3$ monolayer \cite{camerano25,camerano24}. In the octahedral environment $O_h$, the Jahn–Teller distortion \cite{jahn37} lowers the symmetry from $O_h$ to the trigonal point group $D_{3d}$, i.e. splitting the low-energy manifold in a singlet $e_{out}\rightarrow a_{1g}$ and a doublet $d_{in}\rightarrow e^\prime_{g}$. A spontaneous symmetry breaking could completely eliminate orbital degeneracy by further splitting the $d_{in}\rightarrow e^\prime_{g}$ manifold into $e^\prime_{g,1}$ and $e^\prime_{g,2}$. First-principles DFT+U (U = 3.2 eV) calculations showed that the ground state of the pristine VCl$_3$ monolayer with $V^{3+}$ $(3d^2)$ configuration could be stabilized in the metallic phase $a_{1g}e^\prime_{g,1}$ and the insulating phase $e^\prime_{g,1}e^\prime_{g,2}$, both giving rise to magneto-orbital ordering \cite{camerano25}. In the present case, as shown in figure \ref{ZZ}(h), ground state of zero-entropy VCl$_3$ monolayer exhibits the metallic phase $a_{1g}e^\prime_{g,1}$ while the ground state of LRO-entropic V$_{0.5}$(TiCrFeCo)$_{0.5}$Cl$_3$ monolayer displays a small gap spin-gapless semiconducting phase $e^\prime_{g,1}e^\prime_{g,2}$. The entropy-driven phase transition from half-metallic to DSGS phase is in good agreement with the symmetry-identified half-metallic and insulating ground states.

In addition, unlike the zero-entropy VCl$_3$ monolayer, a contribution of Cl-p orbitals is enhanced in the LRO-entropic [ZZ-ttt] monolayer, as shown figure \textbf{S4}(f), leading to a hybridization between spin-up d-orbitals of TM atoms and spin-down p-orbitals of Cl atoms. Such p-d hybridization has also been predicted in Mn- and Fe-doped VCl$_3$ monolayers \cite{ouettar23}. The p-d hybridization can further be validated by inspecting a reduction in magnetic moments of V-3d orbitals. In the zero-entropy VCl$_3$ monolayer, our first principle calculations show that each V$^{3+}$ [$3d^2$] atom contributes a magnetic moment of $m_B=+1.918$ $\mu_B$ while each Cl$^-$ atom contributes $m_B\approx-0.03$ $\mu_B$, as shown in table 1 of Supplementary Information. In the LRO-entropic monolayer [ZZ-ttt], on the other hand, while two of V atoms contribute $m_B\approx+1.90$ $\mu_B$, the magnetic moment of the other two V atoms reduces to $m_B\approx+1.80$ $\mu_B$ and that of Cl atoms varies from $m_B\approx-0.001$ $\mu_B$ to $m_B\approx-0.03$ $\mu_B$.

The diverse nature of magnetic coupling between the TM-3d orbitals and the surrounding Cl-p orbitals also affects the magnetic moments of TM-atoms, leading to a ferromagnetic ground state with a total magnetic moment of $m_B\approx+10.9729$ $\mu_B$, much smaller than the magnetic moment of zero-entropy case ($\approx16$ $\mu_B$). Interestingly, unlike an antiferromagnetic coupling Cl$\downarrow$-TM$\uparrow$-Cl$\downarrow$ between the local magnetic moments of Cl atoms and TM atoms (V, Ti, Co, and Cr), a negative value of the local magnetic moment of the Fe cation indicates a ferromagnetic coupling Cl$\downarrow$-Fe$\downarrow$-Cl$\downarrow$ of the Fe atom with neighboring Cl atoms. This shows that entropy engineering could also be a promising design concept to control magnetic textures in real space. It has been predicted through first-principle calculations that the electronic and magnetic properties of VCl$_3$ monolayer can be tuned through single-atom substitutional doping with 3d TM atoms (Sc, Ti, Cr, Mn, Fe)  \cite{ouettar23}. However, as shown in Table 2 of the Supplementary Information, it is crucial to note that TM atoms doped in a LRO-entropic environment exhibit characteristics that are distinct from the doping of individual TM atoms in zero-entropy VCl$_3$. Further details about the magnetic moments of individual TM atoms and local spin textures for various LRO-entropic configurations are summarized in the Supplementary Information.

The renormalization of electronic and magnetic properties of LRO-entropic V$_{0.5}$(TiCrFeCo)$_{0.5}$Cl$_3$ monolayer [ZZ-ttt] shows that an interplay between electrons localization and de-localization emerges as a trademark of entropy engineering. While band flattening suggests that the decrease in the bandwidth is a direct consequence of entropy manipulation that reduces the inter-orbital overlap, a considerable contribution from Cl atoms and the reduced magnetic moments of TM atoms constituting low-energy bands indicate a hybridization between the d-orbitals of TM atoms and the p-orbitals of the surrounding Cl atoms. It shows that the LRO-entropic monolayer [ZZ-ttt] is a special case in the sense that nearest-neighbor hopping on the zigzag chains made of V-atoms predominantly constitutes low-energy Dirac bands, featuring an intrinsic topological characteristic of honeycomb structures, while the entropy level raised by doped M$^\prime$ atoms reduces the bandwidth of Dirac bands. The interplay between electrons localization and de-localization is also evident from the electronic and magnetic properties of other LRO-entropic configurations. For example, in the Mix-t$^\prime$t$^\prime$t configuration with a total magnetic moment of $m_B\approx+3.286$ $\mu_B$, an interplay between localization and de-localization leads to the band dispersion that exhibits a nodal-line semi-metallic character between nearly flat spin-up and spin-down bands, as shown in figure \textbf{S2}(b). On the other hand, in the SL-t$^\prime$t$^\prime$t$^\prime$ configuration with a total magnetic moment of $m_B\approx+12.143$ $\mu_B$, localization dominates over de-localization leading to a band dispersion exhibiting ferromagnetic insulating behavior, as shown in figure \textbf{S2}(a). This shows that entropy engineering is an effective mechanism for controlling the bandstructure through an interplay between electrons localization and de-localization.

\section{Conclusion and outlook}
Bandstructure renormalization and energy splitting of low-lying orbitals through entropy engineering are attributed to structural symmetry breaking caused by TM dopants in transition-metal trihalides \cite{tomar19,ouettar23}. In addition, apart from the level of entropy and the nature of symmetry breaking, the electronic and magnetic properties of entropic materials rely on what elements are contributing to the low-energy bands and how their low-lying orbitals are influenced by the nearest-neighbor environments. In order to understand the impact of the nearest-neighbor environment on the electronic and magnetic properties, we systematically changed the sublattice site position of doped TM atoms with parent V atoms while keeping the level of entropy fixed. Our first-principle calculations show that a small change in the nearest-neighbor environment could lead to a significant change in electronic and magnetic properties.

The bandstructure renormalization in all four configurations of the long-range ordered V$_{0.5}$(TiCrFeCo)$_{0.5}$Cl$_3$ monolayer is evidently an entropy-driven effect, directly related to an increase in the number of dopants (entropy) and the nearest-neighbor hopping (local environment). However, the ZZ-ttt configuration is special in the sense that it exhibits quantum phases that feature an intrinsic topological character of honeycomb lattice structures. A much relevant and needed question to be addressed could be which of these four long-range ordered configurations is the most favorable to achieve thermodynamic stabilization. In biatomic systems, the SL-t$^\prime$t$^\prime$t$^\prime$ configuration could be more favorable due to structural chirality, as demonstrated recently in the Mn/V-doped CrI$_3$ crystals, i.e., CrMnI$_6$ \cite{zhang20-CrMnI6} and CrVI$_6$ \cite{li24}, which have been experimentally synthesized based on DFT predictions. However, the ZZ-ttt configuration could be more favorable in multi-component systems because of the intrinsic momentum-space topology and its affects on phonon-dispersion in honeycomb structures. A further study would be required to comprehensively address this question based on the intrinsic topological character of the honeycomb lattice structures, the nearest-neighbor environment and the electronic configuration of doped TM atoms quantifying the hopping amplitudes, and a thorough investigation of the thermal history and phonon dispersions for accessing stability.

A robust QAH effect is a unique feature of SGSs \cite{nadeem20, nadeem24}, i.e., SOI opens a nontrivial energy gap by lifting the only twofold degeneracy of the Dirac point(s) in the bulk band dispersion that corresponds to the fully spin-polarized chiral modes in the edge state spectrum. In 2D magnets, the electronic transition from ferromagnetic DHM to a ferromagnetic DSGS and thus a robust QAH effect could open a new route to revolutionize low-energy and high-performance electronic and spintronic technologies, such as giant magnetoresistance devices \cite{de83,Igor04,nadeem24} overcoming the Schmidt obstacle \cite{SFET-Schmidt,Xu15} and topological switching devices \cite{ezawa13,zhang17,xu19-TTFET,nadeem24} overcoming the Boltzmann tyranny \cite{nadeem21,fuhrer21,nadeem22,banerjee22,nadeem24,weber24}. Although the entropy-driven QAH phase is robust, i.e., chiral edge states are completely disentangled from the bulk modes, the QAH gap in entropic VCl$_3$ is smaller than the QAH gap in zero-entropy VCl$_3$. This reduction in the nontrivial bandgap is consistent with the role of SOI and the Berry curvature which capitalize on their maximal strength at valley points. It motivates further research to find suitable TM atoms that could reduce the crystal-field effect and enhance the nontrivial bandgap via blue shift at Dirac points, leading to a room-temperature realization of topological quantum device applications.

VCl$_3$ is a rich system that hosts emergent quasiparticles such as polarons \cite{mastrippolito23}, skyrmions \cite{tran24}, and magneto-orbitons via magneto-orbital coupling \cite{camerano25}. It has also been demonstrated, via first-principle calculations combined with nonequilibrium Green's function, that a magnetic tunnel junction VCl$_3$/CoBr$_3$/VCl$_3$ made of ferromagnetic half-metallic VCl$_3$ exhibits a perfect spin filtering effect and a high tunnel magnetoresistance ratio ($\approx 4.5\times10^{12}$ $\%$) \cite{feng20}. Moreover, a coexistence of ferroelectricity and magnetism has recently been observed in a VCl$_3$ monolayer selectively grown on an NbSe$_2$ substrate \cite{deng25}. In addition, like several other 2D materials with honeycomb lattice structures such as 2D-Xenes \cite{pan14,zhou17} and van der Waals (vdW) heterostructures constructed by transition metal dichalcogenides and transition metal halides \cite{zhang20}, VCl$_3$ also hosts valley-polarized QAH states due to the coexistence of QAH and QVH effects \cite{li18,liu18}. In zigzag nanoribbons of honeycomb lattice structures, because of an intrinsic topology and a nontrivial Berry curvature intertwined with valley DOF, spin-valley coupling and the valley-polarized states promise a wide range of applications in spin electronics and valley electronics. Considering the proposed proof-of-concept robustness of the QAH effect via entropy engineering, it would be interesting to explore how entropy engineering could further enhance the splitting of the valley polarization state to realize the robust valley-polarized QAH effect in a wider range of material systems.

The proposed concept for materials design and discovery of DSGSs through entropy engineering, employed here for VCl$_3$ monolayer, could also be extended for other 2D DHMs and 2D Dirac materials. For instance, like VCl$_3$ monolayer, intrinsic magnetism \cite{tomar19}, spin gapless electronic structure, and the corresponding SOI-induced QAH effect could be renormalized in various other monolayers of transition metal trihalides such as OsCl$_3$\cite{OsCl}, RuX$_3$ (X:Br,Cl,I)\cite{RuI}, MnX$_3$ (X:F,Cl,Br,I)\cite{MnX}, NiCl$_3$\cite{NiCl}, PtCl$_3$\cite{PtCl-You19}, PdCl$_3$\cite{wang18-pdcl}, MBr$_3$ (M:Pd,Pt)\cite{you19-PRA}, MBr$_3$ (M:V,Fe,Ni,Pd) \cite{sun20}, FeX$_3$ (X:Cl,Br,I)\cite{li19}, and VX$_3$ (X:Cl,I)\cite{he16,zhou16,feng20,ouettar23}. Similarly, the proposed design concept could also be employed for enhancing the robustness of QAH phase in magnetic-doped topological insulators, i.e., dissipative bulk channels, and thus backscattering could potentially be removed by increasing the configurational entropy.

Recently, single crystalline VCl$_3$ has been grown \cite{mastrippolito23} using the self-selecting vapor growth (SSVG) technique and structural characterization of the as-grown crystals is presented by energy-dispersive X-ray (EDX) analysis and x-ray photoemission spectroscopy (XPS). In addition, a structural phase transition at T = 103 K and an antiferromagnetic transition at $T_N = 21.8$ K are observed through temperature-dependent heat capacity measurements using the Evercool II Quantum Design Physical Property Measurement System (PPMS). The nature of magnetism and the role of symmetry breaking is also demonstrated through experimental observation and first-principles calculations in a recently synthesized VCl$_3$-NbSe$_2$ heterojunction \cite{deng25}. In an epitaxially grown VCl$_3$ monolayer on an NbSe$_2$ substrate, vdW interfacial Cl-Se interactions break both in-plane $C_3$ rotational and out-of-pane inversion symmetries, inducing in-plane ferroelectricity accompanied by antiferromagnetic order with canted magnetic moments. Scanning tunneling microscopy (STM) and scanning tunneling spectroscopy (STS) provide consistent evidence that the VCl$_3$ monolayer hosts in-plane ferroelectricity while a bistriped antiferromagnetic order ($T_N = 16$ K) with an easy-plane \textit{xz} is confirmed by vibrating sample magnetometry (VSM). In addition, based on DFT predictions, biatomic transition-metal trihalide CrVI$_6$ \cite{li24} has recently been experimentally synthesized. The experimental progress and confirmation of emergent functionalities in vanadium-based trihalides motivate the synthesis of an entropic VCl$_3$ monolayer where entropy engineering could play an elegant role in the tailoring of crystalline symmetries and the corresponding emergent quantum phases.

Compared to single-atom substitutional doping in the VCl$_3$ monolayer \cite{ouettar23}, thermodynamic phase stability is more favorable in an entropic VCl$_3$ monolayer. The configurational entropy is correlated with the phase transition and phase stabilization. That is, a phase transition from a thermodynamically unstable phase to a thermodynamically stable phase is associated with entropy-driven phase stabilization \cite{jiang21}. The general concept of entropy-dominated phase stabilization is based on the reduction in the Gibbs free energy $\Delta G$ ($\Delta G=\Delta H- T\Delta S$), predominantly by increasing the entropy $\Delta S$ of the system. This can be done by introducing multiple elements with random distribution at the same lattice sites.

The topological and magnetic properties can change with temperature. Although, the origin of the nontrivial topology in the entropic V$_{0.5}$(TiCrFeCo)$_{0.5}$Cl$_3$ monolayer is intertwined with the intrinsic topological character of honeycomb lattice structures \cite{nadeem22}, the temperature dependence of the band topology in the entropic V$_{0.5}$(TiCrFeCo)$_{0.5}$Cl$_3$ monolayer can change from that for the pristine VCl$_3$ structure, mainly due to entropy-driven effects on phonon dispersion and an intimate coupling between phonons and various emergent electronic and magnetic quasiparticles. On the same grounds, the Curie temperature and its temperature dependence in the V$_{0.5}$(TiCrFeCo)$_{0.5}$Cl$_3$ monolayer could be different from those in the VCl$_3$ monolayer. Furthermore, as shown in the supplementary information, first-principles analysis of the temperature dependence of the Curie temperature becomes more demanding in these entropic configurations, in which magnetic moments of TM atoms are sensitively dependent on the nearest-neighbor environment. A comprehensive analysis on the temperature dependence of the band topology and Curie temperature in the V$_{0.5}$(TiCrFeCo)$_{0.5}$Cl$_3$ monolayer is beyond the scope of the present study, which presents a proof-of-concept for tailoring a robust QAH effect via entropy engineering. Considering the fundamental importance of the performance stability of materials in high-temperature environments, a detailed analysis on the temperature dependence of topological and magnetic properties in entropic V$_{0.5}$(TiCrFeCo)$_{0.5}$Cl$_3$ monolayer would pave the way for practical applications.

\textbf{Note Added:} In the Mix-t$^\prime$t$^\prime$t LRO-entropic configuration, which exhibits nodal-line semi-metallic character between nearly flat spin-up and spin-down bands, QAH phase could also be realized with a k-dependent Rashba SOI \cite{Bychkov-SOI84,manchon15,anandan23}, distinct from the k-independent Rashba SOI \cite{rashba09} specialized in honeycomb structures. Such an intertwining between Rashba SOI and chirality provides a shared platform for non-reciprocity in Chern magnets and superconductors \cite{nadeem23-NP,nadeem23}.

\begin{acknowledgments}
This research was supported by the Australian Research Council (ARC) through the Discovery Project No. DP230102221 and Linkage Project No. LP220100085.
\end{acknowledgments}

%\nocite{*}
\bibliography{apssamp}

\providecommand{\noopsort}[1]{}\providecommand{\singleletter}[1]{#1}%
\begin{thebibliography}{74}%
\makeatletter
\providecommand \@ifxundefined [1]{%
 \@ifx{#1\undefined}
}%
\providecommand \@ifnum [1]{%
 \ifnum #1\expandafter \@firstoftwo
 \else \expandafter \@secondoftwo
 \fi
}%
\providecommand \@ifx [1]{%
 \ifx #1\expandafter \@firstoftwo
 \else \expandafter \@secondoftwo
 \fi
}%
\providecommand \natexlab [1]{#1}%
\providecommand \enquote  [1]{``#1''}%
\providecommand \bibnamefont  [1]{#1}%
\providecommand \bibfnamefont [1]{#1}%
\providecommand \citenamefont [1]{#1}%
\providecommand \href@noop [0]{\@secondoftwo}%
\providecommand \href [0]{\begingroup \@sanitize@url \@href}%
\providecommand \@href[1]{\@@startlink{#1}\@@href}%
\providecommand \@@href[1]{\endgroup#1\@@endlink}%
\providecommand \@sanitize@url [0]{\catcode `\\12\catcode `\$12\catcode `\&12\catcode `\#12\catcode `\^12\catcode `\_12\catcode `\%12\relax}%
\providecommand \@@startlink[1]{}%
\providecommand \@@endlink[0]{}%
\providecommand \url  [0]{\begingroup\@sanitize@url \@url }%
\providecommand \@url [1]{\endgroup\@href {#1}{\urlprefix }}%
\providecommand \urlprefix  [0]{URL }%
\providecommand \Eprint [0]{\href }%
\providecommand \doibase [0]{https://doi.org/}%
\providecommand \selectlanguage [0]{\@gobble}%
\providecommand \bibinfo  [0]{\@secondoftwo}%
\providecommand \bibfield  [0]{\@secondoftwo}%
\providecommand \translation [1]{[#1]}%
\providecommand \BibitemOpen [0]{}%
\providecommand \bibitemStop [0]{}%
\providecommand \bibitemNoStop [0]{.\EOS\space}%
\providecommand \EOS [0]{\spacefactor3000\relax}%
\providecommand \BibitemShut  [1]{\csname bibitem#1\endcsname}%
\let\auto@bib@innerbib\@empty
%</preamble>
\bibitem [{\citenamefont {Wang}(2008)}]{wang08}%
  \BibitemOpen
  \bibfield  {author} {\bibinfo {author} {\bibfnamefont {X.~L.}\ \bibnamefont {Wang}},\ }\bibfield  {title} {\bibinfo {title} {Proposal for a new class of materials: spin gapless semiconductors},\ }\href@noop {} {\bibfield  {journal} {\bibinfo  {journal} {Physical review letters}\ }\textbf {\bibinfo {volume} {100}},\ \bibinfo {pages} {156404} (\bibinfo {year} {2008})}\BibitemShut {NoStop}%
\bibitem [{\citenamefont {Nadeem}\ and\ \citenamefont {Wang}(2024)}]{nadeem24}%
  \BibitemOpen
  \bibfield  {author} {\bibinfo {author} {\bibfnamefont {M.}~\bibnamefont {Nadeem}}\ and\ \bibinfo {author} {\bibfnamefont {X.}~\bibnamefont {Wang}},\ }\bibfield  {title} {\bibinfo {title} {Spin gapless quantum materials and devices},\ }\href@noop {} {\bibfield  {journal} {\bibinfo  {journal} {Advanced Materials}\ }\textbf {\bibinfo {volume} {36}},\ \bibinfo {pages} {2402503} (\bibinfo {year} {2024})}\BibitemShut {NoStop}%
\bibitem [{\citenamefont {Kossut}\ and\ \citenamefont {Dobrowolski}(1993)}]{kossut93}%
  \BibitemOpen
  \bibfield  {author} {\bibinfo {author} {\bibfnamefont {J.}~\bibnamefont {Kossut}}\ and\ \bibinfo {author} {\bibfnamefont {W.}~\bibnamefont {Dobrowolski}},\ }\bibfield  {title} {\bibinfo {title} {Diluted magnetic semiconductors},\ }\href@noop {} {\bibfield  {journal} {\bibinfo  {journal} {Handbook of Magnetic Materials}\ }\textbf {\bibinfo {volume} {7}},\ \bibinfo {pages} {231} (\bibinfo {year} {1993})}\BibitemShut {NoStop}%
\bibitem [{\citenamefont {De~Groot}\ \emph {et~al.}(1983)\citenamefont {De~Groot}, \citenamefont {Mueller}, \citenamefont {van Engen},\ and\ \citenamefont {Buschow}}]{de83}%
  \BibitemOpen
  \bibfield  {author} {\bibinfo {author} {\bibfnamefont {R.}~\bibnamefont {De~Groot}}, \bibinfo {author} {\bibfnamefont {F.}~\bibnamefont {Mueller}}, \bibinfo {author} {\bibfnamefont {P.~v.}\ \bibnamefont {van Engen}},\ and\ \bibinfo {author} {\bibfnamefont {K.}~\bibnamefont {Buschow}},\ }\bibfield  {title} {\bibinfo {title} {New class of materials: half-metallic ferromagnets},\ }\href@noop {} {\bibfield  {journal} {\bibinfo  {journal} {Physical review letters}\ }\textbf {\bibinfo {volume} {50}},\ \bibinfo {pages} {2024} (\bibinfo {year} {1983})}\BibitemShut {NoStop}%
\bibitem [{\citenamefont {Van~Leuken}\ and\ \citenamefont {De~Groot}(1995)}]{van95}%
  \BibitemOpen
  \bibfield  {author} {\bibinfo {author} {\bibfnamefont {H.}~\bibnamefont {Van~Leuken}}\ and\ \bibinfo {author} {\bibfnamefont {R.}~\bibnamefont {De~Groot}},\ }\bibfield  {title} {\bibinfo {title} {Half-metallic antiferromagnets},\ }\href@noop {} {\bibfield  {journal} {\bibinfo  {journal} {Physical review letters}\ }\textbf {\bibinfo {volume} {74}},\ \bibinfo {pages} {1171} (\bibinfo {year} {1995})}\BibitemShut {NoStop}%
\bibitem [{\citenamefont {Haldane}(1988)}]{haldane88}%
  \BibitemOpen
  \bibfield  {author} {\bibinfo {author} {\bibfnamefont {F.~D.~M.}\ \bibnamefont {Haldane}},\ }\bibfield  {title} {\bibinfo {title} {Model for a quantum hall effect without landau levels: Condensed-matter realization of the" parity anomaly"},\ }\href@noop {} {\bibfield  {journal} {\bibinfo  {journal} {Physical review letters}\ }\textbf {\bibinfo {volume} {61}},\ \bibinfo {pages} {2015} (\bibinfo {year} {1988})}\BibitemShut {NoStop}%
\bibitem [{\citenamefont {Nadeem}\ \emph {et~al.}(2020)\citenamefont {Nadeem}, \citenamefont {Hamilton}, \citenamefont {Fuhrer},\ and\ \citenamefont {Wang}}]{nadeem20}%
  \BibitemOpen
  \bibfield  {author} {\bibinfo {author} {\bibfnamefont {M.}~\bibnamefont {Nadeem}}, \bibinfo {author} {\bibfnamefont {A.~R.}\ \bibnamefont {Hamilton}}, \bibinfo {author} {\bibfnamefont {M.~S.}\ \bibnamefont {Fuhrer}},\ and\ \bibinfo {author} {\bibfnamefont {X.}~\bibnamefont {Wang}},\ }\bibfield  {title} {\bibinfo {title} {Quantum anomalous hall effect in magnetic doped topological insulators and ferromagnetic spin-gapless semiconductors—a perspective review},\ }\href@noop {} {\bibfield  {journal} {\bibinfo  {journal} {Small}\ }\textbf {\bibinfo {volume} {16}},\ \bibinfo {pages} {1904322} (\bibinfo {year} {2020})}\BibitemShut {NoStop}%
\bibitem [{\citenamefont {Wang}(2017)}]{wang17}%
  \BibitemOpen
  \bibfield  {author} {\bibinfo {author} {\bibfnamefont {X.-L.}\ \bibnamefont {Wang}},\ }\bibfield  {title} {\bibinfo {title} {Dirac spin-gapless semiconductors: promising platforms for massless and dissipationless spintronics and new (quantum) anomalous spin hall effects},\ }\href@noop {} {\bibfield  {journal} {\bibinfo  {journal} {National Science Review}\ }\textbf {\bibinfo {volume} {4}},\ \bibinfo {pages} {252} (\bibinfo {year} {2017})}\BibitemShut {NoStop}%
\bibitem [{\citenamefont {Ding}\ \emph {et~al.}(2022)\citenamefont {Ding}, \citenamefont {Wang}, \citenamefont {Chen}, \citenamefont {Zhang},\ and\ \citenamefont {Wang}}]{ding22}%
  \BibitemOpen
  \bibfield  {author} {\bibinfo {author} {\bibfnamefont {G.}~\bibnamefont {Ding}}, \bibinfo {author} {\bibfnamefont {J.}~\bibnamefont {Wang}}, \bibinfo {author} {\bibfnamefont {H.}~\bibnamefont {Chen}}, \bibinfo {author} {\bibfnamefont {X.}~\bibnamefont {Zhang}},\ and\ \bibinfo {author} {\bibfnamefont {X.}~\bibnamefont {Wang}},\ }\bibfield  {title} {\bibinfo {title} {Investigation of nodal line spin-gapless semiconductors using first-principles calculations},\ }\href@noop {} {\bibfield  {journal} {\bibinfo  {journal} {Journal of Materials Chemistry C}\ }\textbf {\bibinfo {volume} {10}},\ \bibinfo {pages} {6530} (\bibinfo {year} {2022})}\BibitemShut {NoStop}%
\bibitem [{\citenamefont {Şaş{\i}oğlu}\ \emph {et~al.}(2019)\citenamefont {Şaş{\i}oğlu}, \citenamefont {Blügel},\ and\ \citenamefont {Mertig}}]{SGS-D19}%
  \BibitemOpen
  \bibfield  {author} {\bibinfo {author} {\bibfnamefont {E.}~\bibnamefont {Şaş{\i}oğlu}}, \bibinfo {author} {\bibfnamefont {S.}~\bibnamefont {Blügel}},\ and\ \bibinfo {author} {\bibfnamefont {I.}~\bibnamefont {Mertig}},\ }\bibfield  {title} {\bibinfo {title} {Proposal for reconfigurable magnetic tunnel diode and transistor},\ }\href@noop {} {\bibfield  {journal} {\bibinfo  {journal} {ACS Applied Electronic Materials}\ }\textbf {\bibinfo {volume} {1}},\ \bibinfo {pages} {1552} (\bibinfo {year} {2019})}\BibitemShut {NoStop}%
\bibitem [{\citenamefont {{\c{S}}a{\c{s}}{\i}o{\u{g}}lu}\ \emph {et~al.}(2020)\citenamefont {{\c{S}}a{\c{s}}{\i}o{\u{g}}lu}, \citenamefont {Aull}, \citenamefont {Kutschabsky}, \citenamefont {Bl{\"u}gel},\ and\ \citenamefont {Mertig}}]{SGS-D20}%
  \BibitemOpen
  \bibfield  {author} {\bibinfo {author} {\bibfnamefont {E.}~\bibnamefont {{\c{S}}a{\c{s}}{\i}o{\u{g}}lu}}, \bibinfo {author} {\bibfnamefont {T.}~\bibnamefont {Aull}}, \bibinfo {author} {\bibfnamefont {D.}~\bibnamefont {Kutschabsky}}, \bibinfo {author} {\bibfnamefont {S.}~\bibnamefont {Bl{\"u}gel}},\ and\ \bibinfo {author} {\bibfnamefont {I.}~\bibnamefont {Mertig}},\ }\bibfield  {title} {\bibinfo {title} {Half-metal--spin-gapless-semiconductor junctions as a route to the ideal diode},\ }\href@noop {} {\bibfield  {journal} {\bibinfo  {journal} {Physical review applied}\ }\textbf {\bibinfo {volume} {14}},\ \bibinfo {pages} {014082} (\bibinfo {year} {2020})}\BibitemShut {NoStop}%
\bibitem [{\citenamefont {Aull}\ \emph {et~al.}(2022)\citenamefont {Aull}, \citenamefont {{\c{S}}a{\c{s}}{\i}o{\u{g}}lu}, \citenamefont {Hinsche},\ and\ \citenamefont {Mertig}}]{SGS-D22}%
  \BibitemOpen
  \bibfield  {author} {\bibinfo {author} {\bibfnamefont {T.}~\bibnamefont {Aull}}, \bibinfo {author} {\bibfnamefont {E.}~\bibnamefont {{\c{S}}a{\c{s}}{\i}o{\u{g}}lu}}, \bibinfo {author} {\bibfnamefont {N.}~\bibnamefont {Hinsche}},\ and\ \bibinfo {author} {\bibfnamefont {I.}~\bibnamefont {Mertig}},\ }\bibfield  {title} {\bibinfo {title} {Ab initio study of magnetic tunnel junctions based on half-metallic and spin-gapless semiconducting heusler compounds: Reconfigurable diode and inverse tunnel-magnetoresistance effect},\ }\href@noop {} {\bibfield  {journal} {\bibinfo  {journal} {Physical Review Applied}\ }\textbf {\bibinfo {volume} {18}},\ \bibinfo {pages} {034024} (\bibinfo {year} {2022})}\BibitemShut {NoStop}%
\bibitem [{\citenamefont {Maji}\ and\ \citenamefont {Nath}(2022)}]{Maji22}%
  \BibitemOpen
  \bibfield  {author} {\bibinfo {author} {\bibfnamefont {N.}~\bibnamefont {Maji}}\ and\ \bibinfo {author} {\bibfnamefont {T.~K.}\ \bibnamefont {Nath}},\ }\bibfield  {title} {\bibinfo {title} {Demonstration of reconfigurable magnetic tunnel diode and giant tunnel magnetoresistance in magnetic tunnel junctions made with spin gapless semiconductor and half-metallic heusler alloy},\ }\href@noop {} {\bibfield  {journal} {\bibinfo  {journal} {Applied Physics Letters}\ }\textbf {\bibinfo {volume} {120}} (\bibinfo {year} {2022})}\BibitemShut {NoStop}%
\bibitem [{\citenamefont {Feng}\ \emph {et~al.}(2020)\citenamefont {Feng}, \citenamefont {Wu},\ and\ \citenamefont {Gao}}]{feng20}%
  \BibitemOpen
  \bibfield  {author} {\bibinfo {author} {\bibfnamefont {Y.}~\bibnamefont {Feng}}, \bibinfo {author} {\bibfnamefont {X.}~\bibnamefont {Wu}},\ and\ \bibinfo {author} {\bibfnamefont {G.}~\bibnamefont {Gao}},\ }\bibfield  {title} {\bibinfo {title} {High tunnel magnetoresistance based on 2d dirac spin gapless semiconductor vcl3},\ }\href@noop {} {\bibfield  {journal} {\bibinfo  {journal} {Applied Physics Letters}\ }\textbf {\bibinfo {volume} {116}} (\bibinfo {year} {2020})}\BibitemShut {NoStop}%
\bibitem [{\citenamefont {Murty}\ \emph {et~al.}(2019)\citenamefont {Murty}, \citenamefont {Yeh}, \citenamefont {Ranganathan},\ and\ \citenamefont {Bhattacharjee}}]{murty19-book}%
  \BibitemOpen
  \bibfield  {author} {\bibinfo {author} {\bibfnamefont {B.~S.}\ \bibnamefont {Murty}}, \bibinfo {author} {\bibfnamefont {J.-W.}\ \bibnamefont {Yeh}}, \bibinfo {author} {\bibfnamefont {S.}~\bibnamefont {Ranganathan}},\ and\ \bibinfo {author} {\bibfnamefont {P.}~\bibnamefont {Bhattacharjee}},\ }\href@noop {} {\emph {\bibinfo {title} {High-entropy alloys}}}\ (\bibinfo  {publisher} {Elsevier},\ \bibinfo {year} {2019})\BibitemShut {NoStop}%
\bibitem [{\citenamefont {Sarkar}\ \emph {et~al.}(2019)\citenamefont {Sarkar}, \citenamefont {Wang}, \citenamefont {Schiele}, \citenamefont {Chellali}, \citenamefont {Bhattacharya}, \citenamefont {Wang}, \citenamefont {Brezesinski}, \citenamefont {Hahn}, \citenamefont {Velasco},\ and\ \citenamefont {Breitung}}]{sarkar19-AM}%
  \BibitemOpen
  \bibfield  {author} {\bibinfo {author} {\bibfnamefont {A.}~\bibnamefont {Sarkar}}, \bibinfo {author} {\bibfnamefont {Q.}~\bibnamefont {Wang}}, \bibinfo {author} {\bibfnamefont {A.}~\bibnamefont {Schiele}}, \bibinfo {author} {\bibfnamefont {M.~R.}\ \bibnamefont {Chellali}}, \bibinfo {author} {\bibfnamefont {S.~S.}\ \bibnamefont {Bhattacharya}}, \bibinfo {author} {\bibfnamefont {D.}~\bibnamefont {Wang}}, \bibinfo {author} {\bibfnamefont {T.}~\bibnamefont {Brezesinski}}, \bibinfo {author} {\bibfnamefont {H.}~\bibnamefont {Hahn}}, \bibinfo {author} {\bibfnamefont {L.}~\bibnamefont {Velasco}},\ and\ \bibinfo {author} {\bibfnamefont {B.}~\bibnamefont {Breitung}},\ }\bibfield  {title} {\bibinfo {title} {High-entropy oxides: fundamental aspects and electrochemical properties},\ }\href@noop {} {\bibfield  {journal} {\bibinfo  {journal} {Advanced Materials}\ }\textbf {\bibinfo {volume} {31}},\ \bibinfo {pages} {1806236} (\bibinfo {year} {2019})}\BibitemShut {NoStop}%
\bibitem [{\citenamefont {Hsu}\ \emph {et~al.}(2024)\citenamefont {Hsu}, \citenamefont {Tsai}, \citenamefont {Yeh},\ and\ \citenamefont {Yeh}}]{hsu24-NRC}%
  \BibitemOpen
  \bibfield  {author} {\bibinfo {author} {\bibfnamefont {W.-L.}\ \bibnamefont {Hsu}}, \bibinfo {author} {\bibfnamefont {C.-W.}\ \bibnamefont {Tsai}}, \bibinfo {author} {\bibfnamefont {A.-C.}\ \bibnamefont {Yeh}},\ and\ \bibinfo {author} {\bibfnamefont {J.-W.}\ \bibnamefont {Yeh}},\ }\bibfield  {title} {\bibinfo {title} {Clarifying the four core effects of high-entropy materials},\ }\href@noop {} {\bibfield  {journal} {\bibinfo  {journal} {Nature Reviews Chemistry}\ ,\ \bibinfo {pages} {1}} (\bibinfo {year} {2024})}\BibitemShut {NoStop}%
\bibitem [{\citenamefont {Schweidler}\ \emph {et~al.}(2024)\citenamefont {Schweidler}, \citenamefont {Botros}, \citenamefont {Strauss}, \citenamefont {Wang}, \citenamefont {Ma}, \citenamefont {Velasco}, \citenamefont {Cadilha~Marques}, \citenamefont {Sarkar}, \citenamefont {K{\"u}bel}, \citenamefont {Hahn} \emph {et~al.}}]{schweidler24-NRM}%
  \BibitemOpen
  \bibfield  {author} {\bibinfo {author} {\bibfnamefont {S.}~\bibnamefont {Schweidler}}, \bibinfo {author} {\bibfnamefont {M.}~\bibnamefont {Botros}}, \bibinfo {author} {\bibfnamefont {F.}~\bibnamefont {Strauss}}, \bibinfo {author} {\bibfnamefont {Q.}~\bibnamefont {Wang}}, \bibinfo {author} {\bibfnamefont {Y.}~\bibnamefont {Ma}}, \bibinfo {author} {\bibfnamefont {L.}~\bibnamefont {Velasco}}, \bibinfo {author} {\bibfnamefont {G.}~\bibnamefont {Cadilha~Marques}}, \bibinfo {author} {\bibfnamefont {A.}~\bibnamefont {Sarkar}}, \bibinfo {author} {\bibfnamefont {C.}~\bibnamefont {K{\"u}bel}}, \bibinfo {author} {\bibfnamefont {H.}~\bibnamefont {Hahn}}, \emph {et~al.},\ }\bibfield  {title} {\bibinfo {title} {High-entropy materials for energy and electronic applications},\ }\href@noop {} {\bibfield  {journal} {\bibinfo  {journal} {Nature Reviews Materials}\ }\textbf {\bibinfo {volume} {9}},\ \bibinfo {pages} {266} (\bibinfo {year} {2024})}\BibitemShut {NoStop}%
\bibitem [{\citenamefont {Kresse}\ and\ \citenamefont {Hafner}(1993)}]{kresse93}%
  \BibitemOpen
  \bibfield  {author} {\bibinfo {author} {\bibfnamefont {G.}~\bibnamefont {Kresse}}\ and\ \bibinfo {author} {\bibfnamefont {J.}~\bibnamefont {Hafner}},\ }\bibfield  {title} {\bibinfo {title} {Ab initio molecular dynamics for liquid metals},\ }\href@noop {} {\bibfield  {journal} {\bibinfo  {journal} {Physical review B}\ }\textbf {\bibinfo {volume} {47}},\ \bibinfo {pages} {558} (\bibinfo {year} {1993})}\BibitemShut {NoStop}%
\bibitem [{\citenamefont {Kresse}\ and\ \citenamefont {Joubert}(1999)}]{kresse99}%
  \BibitemOpen
  \bibfield  {author} {\bibinfo {author} {\bibfnamefont {G.}~\bibnamefont {Kresse}}\ and\ \bibinfo {author} {\bibfnamefont {D.}~\bibnamefont {Joubert}},\ }\bibfield  {title} {\bibinfo {title} {From ultrasoft pseudopotentials to the projector augmented-wave method},\ }\href@noop {} {\bibfield  {journal} {\bibinfo  {journal} {Physical review b}\ }\textbf {\bibinfo {volume} {59}},\ \bibinfo {pages} {1758} (\bibinfo {year} {1999})}\BibitemShut {NoStop}%
\bibitem [{\citenamefont {Perdew}\ \emph {et~al.}(1996)\citenamefont {Perdew}, \citenamefont {Burke},\ and\ \citenamefont {Ernzerhof}}]{perdew96}%
  \BibitemOpen
  \bibfield  {author} {\bibinfo {author} {\bibfnamefont {J.~P.}\ \bibnamefont {Perdew}}, \bibinfo {author} {\bibfnamefont {K.}~\bibnamefont {Burke}},\ and\ \bibinfo {author} {\bibfnamefont {M.}~\bibnamefont {Ernzerhof}},\ }\bibfield  {title} {\bibinfo {title} {Generalized gradient approximation made simple},\ }\href@noop {} {\bibfield  {journal} {\bibinfo  {journal} {Physical review letters}\ }\textbf {\bibinfo {volume} {77}},\ \bibinfo {pages} {3865} (\bibinfo {year} {1996})}\BibitemShut {NoStop}%
\bibitem [{\citenamefont {Bl{\"o}chl}(1994)}]{blochl94}%
  \BibitemOpen
  \bibfield  {author} {\bibinfo {author} {\bibfnamefont {P.~E.}\ \bibnamefont {Bl{\"o}chl}},\ }\bibfield  {title} {\bibinfo {title} {Projector augmented-wave method},\ }\href@noop {} {\bibfield  {journal} {\bibinfo  {journal} {Physical review B}\ }\textbf {\bibinfo {volume} {50}},\ \bibinfo {pages} {17953} (\bibinfo {year} {1994})}\BibitemShut {NoStop}%
\bibitem [{\citenamefont {Monkhorst}\ and\ \citenamefont {Pack}(1976)}]{monkhorst76}%
  \BibitemOpen
  \bibfield  {author} {\bibinfo {author} {\bibfnamefont {H.~J.}\ \bibnamefont {Monkhorst}}\ and\ \bibinfo {author} {\bibfnamefont {J.~D.}\ \bibnamefont {Pack}},\ }\bibfield  {title} {\bibinfo {title} {Special points for brillouin-zone integrations},\ }\href@noop {} {\bibfield  {journal} {\bibinfo  {journal} {Physical review B}\ }\textbf {\bibinfo {volume} {13}},\ \bibinfo {pages} {5188} (\bibinfo {year} {1976})}\BibitemShut {NoStop}%
\bibitem [{\citenamefont {Zheng}(2022)}]{QZheng}%
  \BibitemOpen
  \bibfield  {author} {\bibinfo {author} {\bibfnamefont {Q.}~\bibnamefont {Zheng}},\ }\bibfield  {title} {\bibinfo {title} {Pyvaspwfc},\ }\href@noop {} {\bibfield  {journal} {\bibinfo  {journal} {https://github.com/QijingZheng/ VaspBandUnfolding}\ } (\bibinfo {year} {2022})}\BibitemShut {NoStop}%
\bibitem [{\citenamefont {Wang}\ \emph {et~al.}(2021)\citenamefont {Wang}, \citenamefont {Xu}, \citenamefont {Liu}, \citenamefont {Tang},\ and\ \citenamefont {Geng}}]{wang21}%
  \BibitemOpen
  \bibfield  {author} {\bibinfo {author} {\bibfnamefont {V.}~\bibnamefont {Wang}}, \bibinfo {author} {\bibfnamefont {N.}~\bibnamefont {Xu}}, \bibinfo {author} {\bibfnamefont {J.-C.}\ \bibnamefont {Liu}}, \bibinfo {author} {\bibfnamefont {G.}~\bibnamefont {Tang}},\ and\ \bibinfo {author} {\bibfnamefont {W.-T.}\ \bibnamefont {Geng}},\ }\bibfield  {title} {\bibinfo {title} {Vaspkit: A user-friendly interface facilitating high-throughput computing and analysis using vasp code},\ }\href@noop {} {\bibfield  {journal} {\bibinfo  {journal} {Computer Physics Communications}\ }\textbf {\bibinfo {volume} {267}},\ \bibinfo {pages} {108033} (\bibinfo {year} {2021})}\BibitemShut {NoStop}%
\bibitem [{\citenamefont {Wu}\ \emph {et~al.}(2018)\citenamefont {Wu}, \citenamefont {Zhang}, \citenamefont {Song}, \citenamefont {Troyer},\ and\ \citenamefont {Soluyanov}}]{wu18}%
  \BibitemOpen
  \bibfield  {author} {\bibinfo {author} {\bibfnamefont {Q.}~\bibnamefont {Wu}}, \bibinfo {author} {\bibfnamefont {S.}~\bibnamefont {Zhang}}, \bibinfo {author} {\bibfnamefont {H.-F.}\ \bibnamefont {Song}}, \bibinfo {author} {\bibfnamefont {M.}~\bibnamefont {Troyer}},\ and\ \bibinfo {author} {\bibfnamefont {A.~A.}\ \bibnamefont {Soluyanov}},\ }\bibfield  {title} {\bibinfo {title} {Wanniertools: An open-source software package for novel topological materials},\ }\href@noop {} {\bibfield  {journal} {\bibinfo  {journal} {Computer Physics Communications}\ }\textbf {\bibinfo {volume} {224}},\ \bibinfo {pages} {405} (\bibinfo {year} {2018})}\BibitemShut {NoStop}%
\bibitem [{\citenamefont {Aamlid}\ \emph {et~al.}(2023)\citenamefont {Aamlid}, \citenamefont {Oudah}, \citenamefont {Rottler},\ and\ \citenamefont {Hallas}}]{aamlid23}%
  \BibitemOpen
  \bibfield  {author} {\bibinfo {author} {\bibfnamefont {S.~S.}\ \bibnamefont {Aamlid}}, \bibinfo {author} {\bibfnamefont {M.}~\bibnamefont {Oudah}}, \bibinfo {author} {\bibfnamefont {J.}~\bibnamefont {Rottler}},\ and\ \bibinfo {author} {\bibfnamefont {A.~M.}\ \bibnamefont {Hallas}},\ }\bibfield  {title} {\bibinfo {title} {Understanding the role of entropy in high entropy oxides},\ }\href@noop {} {\bibfield  {journal} {\bibinfo  {journal} {Journal of the American Chemical Society}\ }\textbf {\bibinfo {volume} {145}},\ \bibinfo {pages} {5991} (\bibinfo {year} {2023})}\BibitemShut {NoStop}%
\bibitem [{\citenamefont {He}\ \emph {et~al.}(2016)\citenamefont {He}, \citenamefont {Ma}, \citenamefont {Lyu},\ and\ \citenamefont {Nachtigall}}]{he16}%
  \BibitemOpen
  \bibfield  {author} {\bibinfo {author} {\bibfnamefont {J.}~\bibnamefont {He}}, \bibinfo {author} {\bibfnamefont {S.}~\bibnamefont {Ma}}, \bibinfo {author} {\bibfnamefont {P.}~\bibnamefont {Lyu}},\ and\ \bibinfo {author} {\bibfnamefont {P.}~\bibnamefont {Nachtigall}},\ }\bibfield  {title} {\bibinfo {title} {Unusual dirac half-metallicity with intrinsic ferromagnetism in vanadium trihalide monolayers},\ }\href@noop {} {\bibfield  {journal} {\bibinfo  {journal} {Journal of Materials Chemistry C}\ }\textbf {\bibinfo {volume} {4}},\ \bibinfo {pages} {2518} (\bibinfo {year} {2016})}\BibitemShut {NoStop}%
\bibitem [{\citenamefont {Zhou}\ \emph {et~al.}(2016)\citenamefont {Zhou}, \citenamefont {Lu}, \citenamefont {Zu},\ and\ \citenamefont {Gao}}]{zhou16}%
  \BibitemOpen
  \bibfield  {author} {\bibinfo {author} {\bibfnamefont {Y.}~\bibnamefont {Zhou}}, \bibinfo {author} {\bibfnamefont {H.}~\bibnamefont {Lu}}, \bibinfo {author} {\bibfnamefont {X.}~\bibnamefont {Zu}},\ and\ \bibinfo {author} {\bibfnamefont {F.}~\bibnamefont {Gao}},\ }\bibfield  {title} {\bibinfo {title} {Evidencing the existence of exciting half-metallicity in two-dimensional ticl3 and vcl3 sheets},\ }\href@noop {} {\bibfield  {journal} {\bibinfo  {journal} {Scientific reports}\ }\textbf {\bibinfo {volume} {6}},\ \bibinfo {pages} {19407} (\bibinfo {year} {2016})}\BibitemShut {NoStop}%
\bibitem [{\citenamefont {Ouettar}\ \emph {et~al.}(2023)\citenamefont {Ouettar}, \citenamefont {Yahi}, \citenamefont {Zanat},\ and\ \citenamefont {Chibani}}]{ouettar23}%
  \BibitemOpen
  \bibfield  {author} {\bibinfo {author} {\bibfnamefont {C.}~\bibnamefont {Ouettar}}, \bibinfo {author} {\bibfnamefont {H.}~\bibnamefont {Yahi}}, \bibinfo {author} {\bibfnamefont {K.}~\bibnamefont {Zanat}},\ and\ \bibinfo {author} {\bibfnamefont {H.}~\bibnamefont {Chibani}},\ }\bibfield  {title} {\bibinfo {title} {Tuned electronic and magnetic properties in 3d transition metal doped vcl3 monolayer: a first-principles study},\ }\href@noop {} {\bibfield  {journal} {\bibinfo  {journal} {Physica Scripta}\ }\textbf {\bibinfo {volume} {98}},\ \bibinfo {pages} {025814} (\bibinfo {year} {2023})}\BibitemShut {NoStop}%
\bibitem [{\citenamefont {Klemm}\ and\ \citenamefont {Krose}(1947)}]{klemm47}%
  \BibitemOpen
  \bibfield  {author} {\bibinfo {author} {\bibfnamefont {W.}~\bibnamefont {Klemm}}\ and\ \bibinfo {author} {\bibfnamefont {E.}~\bibnamefont {Krose}},\ }\bibfield  {title} {\bibinfo {title} {Die kristallstrukturen von sccl3, ticl3 und vcl3},\ }\href@noop {} {\bibfield  {journal} {\bibinfo  {journal} {Zeitschrift f{\"u}r anorganische Chemie}\ }\textbf {\bibinfo {volume} {253}},\ \bibinfo {pages} {218} (\bibinfo {year} {1947})}\BibitemShut {NoStop}%
\bibitem [{\citenamefont {Mastrippolito}\ \emph {et~al.}(2023)\citenamefont {Mastrippolito}, \citenamefont {Camerano}, \citenamefont {{\'S}wiatek}, \citenamefont {{\v{S}}m{\'\i}d}, \citenamefont {Klimczuk}, \citenamefont {Ottaviano},\ and\ \citenamefont {Profeta}}]{mastrippolito23}%
  \BibitemOpen
  \bibfield  {author} {\bibinfo {author} {\bibfnamefont {D.}~\bibnamefont {Mastrippolito}}, \bibinfo {author} {\bibfnamefont {L.}~\bibnamefont {Camerano}}, \bibinfo {author} {\bibfnamefont {H.}~\bibnamefont {{\'S}wiatek}}, \bibinfo {author} {\bibfnamefont {B.}~\bibnamefont {{\v{S}}m{\'\i}d}}, \bibinfo {author} {\bibfnamefont {T.}~\bibnamefont {Klimczuk}}, \bibinfo {author} {\bibfnamefont {L.}~\bibnamefont {Ottaviano}},\ and\ \bibinfo {author} {\bibfnamefont {G.}~\bibnamefont {Profeta}},\ }\bibfield  {title} {\bibinfo {title} {Polaronic and mott insulating phase of layered magnetic vanadium trihalide vcl 3},\ }\href@noop {} {\bibfield  {journal} {\bibinfo  {journal} {Physical Review B}\ }\textbf {\bibinfo {volume} {108}},\ \bibinfo {pages} {045126} (\bibinfo {year} {2023})}\BibitemShut {NoStop}%
\bibitem [{\citenamefont {Deng}\ \emph {et~al.}(2025)\citenamefont {Deng}, \citenamefont {Guo}, \citenamefont {Wen}, \citenamefont {Lu}, \citenamefont {Zhang}, \citenamefont {Cheng}, \citenamefont {Pan}, \citenamefont {Jian}, \citenamefont {Li}, \citenamefont {Wang} \emph {et~al.}}]{deng25}%
  \BibitemOpen
  \bibfield  {author} {\bibinfo {author} {\bibfnamefont {J.}~\bibnamefont {Deng}}, \bibinfo {author} {\bibfnamefont {D.}~\bibnamefont {Guo}}, \bibinfo {author} {\bibfnamefont {Y.}~\bibnamefont {Wen}}, \bibinfo {author} {\bibfnamefont {S.}~\bibnamefont {Lu}}, \bibinfo {author} {\bibfnamefont {H.}~\bibnamefont {Zhang}}, \bibinfo {author} {\bibfnamefont {Z.}~\bibnamefont {Cheng}}, \bibinfo {author} {\bibfnamefont {Z.}~\bibnamefont {Pan}}, \bibinfo {author} {\bibfnamefont {T.}~\bibnamefont {Jian}}, \bibinfo {author} {\bibfnamefont {D.}~\bibnamefont {Li}}, \bibinfo {author} {\bibfnamefont {H.}~\bibnamefont {Wang}}, \emph {et~al.},\ }\bibfield  {title} {\bibinfo {title} {Evidence of ferroelectricity in an antiferromagnetic vanadium trichloride monolayer},\ }\href@noop {} {\bibfield  {journal} {\bibinfo  {journal} {Science Advances}\ }\textbf {\bibinfo {volume} {11}},\ \bibinfo {pages} {eado6538} (\bibinfo {year} {2025})}\BibitemShut {NoStop}%
\bibitem [{\citenamefont {Camerano}\ \emph {et~al.}(2025)\citenamefont {Camerano}, \citenamefont {Fumega}, \citenamefont {Profeta},\ and\ \citenamefont {Lado}}]{camerano25}%
  \BibitemOpen
  \bibfield  {author} {\bibinfo {author} {\bibfnamefont {L.}~\bibnamefont {Camerano}}, \bibinfo {author} {\bibfnamefont {A.~O.}\ \bibnamefont {Fumega}}, \bibinfo {author} {\bibfnamefont {G.}~\bibnamefont {Profeta}},\ and\ \bibinfo {author} {\bibfnamefont {J.~L.}\ \bibnamefont {Lado}},\ }\bibfield  {title} {\bibinfo {title} {Multicomponent magneto-orbital order and magneto-orbitons in monolayer vcl3},\ }\href@noop {} {\bibfield  {journal} {\bibinfo  {journal} {Nano Letters}\ }\textbf {\bibinfo {volume} {25}},\ \bibinfo {pages} {4825} (\bibinfo {year} {2025})}\BibitemShut {NoStop}%
\bibitem [{\citenamefont {Camerano}\ and\ \citenamefont {Profeta}(2024)}]{camerano24}%
  \BibitemOpen
  \bibfield  {author} {\bibinfo {author} {\bibfnamefont {L.}~\bibnamefont {Camerano}}\ and\ \bibinfo {author} {\bibfnamefont {G.}~\bibnamefont {Profeta}},\ }\bibfield  {title} {\bibinfo {title} {Symmetry breaking in vanadium trihalides},\ }\href@noop {} {\bibfield  {journal} {\bibinfo  {journal} {2D Materials}\ }\textbf {\bibinfo {volume} {11}},\ \bibinfo {pages} {025027} (\bibinfo {year} {2024})}\BibitemShut {NoStop}%
\bibitem [{\citenamefont {Tomar}\ \emph {et~al.}(2019)\citenamefont {Tomar}, \citenamefont {Ghosh}, \citenamefont {Mardanya}, \citenamefont {Rastogi}, \citenamefont {Bhadoria}, \citenamefont {Chauhan}, \citenamefont {Agarwal},\ and\ \citenamefont {Bhowmick}}]{tomar19}%
  \BibitemOpen
  \bibfield  {author} {\bibinfo {author} {\bibfnamefont {S.}~\bibnamefont {Tomar}}, \bibinfo {author} {\bibfnamefont {B.}~\bibnamefont {Ghosh}}, \bibinfo {author} {\bibfnamefont {S.}~\bibnamefont {Mardanya}}, \bibinfo {author} {\bibfnamefont {P.}~\bibnamefont {Rastogi}}, \bibinfo {author} {\bibfnamefont {B.}~\bibnamefont {Bhadoria}}, \bibinfo {author} {\bibfnamefont {Y.~S.}\ \bibnamefont {Chauhan}}, \bibinfo {author} {\bibfnamefont {A.}~\bibnamefont {Agarwal}},\ and\ \bibinfo {author} {\bibfnamefont {S.}~\bibnamefont {Bhowmick}},\ }\bibfield  {title} {\bibinfo {title} {Intrinsic magnetism in monolayer transition metal trihalides: A comparative study},\ }\href@noop {} {\bibfield  {journal} {\bibinfo  {journal} {Journal of Magnetism and Magnetic Materials}\ }\textbf {\bibinfo {volume} {489}},\ \bibinfo {pages} {165384} (\bibinfo {year} {2019})}\BibitemShut {NoStop}%
\bibitem [{\citenamefont {Sheng}\ and\ \citenamefont {Nikoli{\'c}}(2017)}]{OsCl}%
  \BibitemOpen
  \bibfield  {author} {\bibinfo {author} {\bibfnamefont {X.-L.}\ \bibnamefont {Sheng}}\ and\ \bibinfo {author} {\bibfnamefont {B.~K.}\ \bibnamefont {Nikoli{\'c}}},\ }\bibfield  {title} {\bibinfo {title} {Monolayer of the 5 d transition metal trichloride oscl 3: A playground for two-dimensional magnetism, room-temperature quantum anomalous hall effect, and topological phase transitions},\ }\href@noop {} {\bibfield  {journal} {\bibinfo  {journal} {Physical Review B}\ }\textbf {\bibinfo {volume} {95}},\ \bibinfo {pages} {201402} (\bibinfo {year} {2017})}\BibitemShut {NoStop}%
\bibitem [{\citenamefont {Li}\ \emph {et~al.}(2018)\citenamefont {Li}, \citenamefont {Liu}, \citenamefont {Wang}, \citenamefont {Wang}, \citenamefont {Xu},\ and\ \citenamefont {Duan}}]{li18}%
  \BibitemOpen
  \bibfield  {author} {\bibinfo {author} {\bibfnamefont {Y.}~\bibnamefont {Li}}, \bibinfo {author} {\bibfnamefont {Y.}~\bibnamefont {Liu}}, \bibinfo {author} {\bibfnamefont {C.}~\bibnamefont {Wang}}, \bibinfo {author} {\bibfnamefont {J.}~\bibnamefont {Wang}}, \bibinfo {author} {\bibfnamefont {Y.}~\bibnamefont {Xu}},\ and\ \bibinfo {author} {\bibfnamefont {W.}~\bibnamefont {Duan}},\ }\bibfield  {title} {\bibinfo {title} {Electrically tunable valleytronics in quantum anomalous hall insulating transition metal trihalides},\ }\href@noop {} {\bibfield  {journal} {\bibinfo  {journal} {Physical Review B}\ }\textbf {\bibinfo {volume} {98}},\ \bibinfo {pages} {201407} (\bibinfo {year} {2018})}\BibitemShut {NoStop}%
\bibitem [{\citenamefont {Liu}\ \emph {et~al.}(2018)\citenamefont {Liu}, \citenamefont {Sun}, \citenamefont {Liu},\ and\ \citenamefont {Meng}}]{liu18}%
  \BibitemOpen
  \bibfield  {author} {\bibinfo {author} {\bibfnamefont {H.}~\bibnamefont {Liu}}, \bibinfo {author} {\bibfnamefont {J.-T.}\ \bibnamefont {Sun}}, \bibinfo {author} {\bibfnamefont {M.}~\bibnamefont {Liu}},\ and\ \bibinfo {author} {\bibfnamefont {S.}~\bibnamefont {Meng}},\ }\bibfield  {title} {\bibinfo {title} {Screening magnetic two-dimensional atomic crystals with nontrivial electronic topology},\ }\href@noop {} {\bibfield  {journal} {\bibinfo  {journal} {The Journal of Physical Chemistry Letters}\ }\textbf {\bibinfo {volume} {9}},\ \bibinfo {pages} {6709} (\bibinfo {year} {2018})}\BibitemShut {NoStop}%
\bibitem [{\citenamefont {Shabbir}\ \emph {et~al.}(2018)\citenamefont {Shabbir}, \citenamefont {Nadeem}, \citenamefont {Dai}, \citenamefont {Fuhrer}, \citenamefont {Xue}, \citenamefont {Wang},\ and\ \citenamefont {Bao}}]{Babar19}%
  \BibitemOpen
  \bibfield  {author} {\bibinfo {author} {\bibfnamefont {B.}~\bibnamefont {Shabbir}}, \bibinfo {author} {\bibfnamefont {M.}~\bibnamefont {Nadeem}}, \bibinfo {author} {\bibfnamefont {Z.}~\bibnamefont {Dai}}, \bibinfo {author} {\bibfnamefont {M.~S.}\ \bibnamefont {Fuhrer}}, \bibinfo {author} {\bibfnamefont {Q.-K.}\ \bibnamefont {Xue}}, \bibinfo {author} {\bibfnamefont {X.}~\bibnamefont {Wang}},\ and\ \bibinfo {author} {\bibfnamefont {Q.}~\bibnamefont {Bao}},\ }\bibfield  {title} {\bibinfo {title} {Long range intrinsic ferromagnetism in two dimensional materials and dissipationless future technologies},\ }\href@noop {} {\bibfield  {journal} {\bibinfo  {journal} {Applied Physics Reviews}\ }\textbf {\bibinfo {volume} {5}} (\bibinfo {year} {2018})}\BibitemShut {NoStop}%
\bibitem [{\citenamefont {Nadeem}\ \emph {et~al.}(2022)\citenamefont {Nadeem}, \citenamefont {Zhang}, \citenamefont {Culcer}, \citenamefont {Hamilton}, \citenamefont {Fuhrer},\ and\ \citenamefont {Wang}}]{nadeem22}%
  \BibitemOpen
  \bibfield  {author} {\bibinfo {author} {\bibfnamefont {M.}~\bibnamefont {Nadeem}}, \bibinfo {author} {\bibfnamefont {C.}~\bibnamefont {Zhang}}, \bibinfo {author} {\bibfnamefont {D.}~\bibnamefont {Culcer}}, \bibinfo {author} {\bibfnamefont {A.~R.}\ \bibnamefont {Hamilton}}, \bibinfo {author} {\bibfnamefont {M.~S.}\ \bibnamefont {Fuhrer}},\ and\ \bibinfo {author} {\bibfnamefont {X.}~\bibnamefont {Wang}},\ }\bibfield  {title} {\bibinfo {title} {Optimizing topological switching in confined 2d-xene nanoribbons via finite-size effects},\ }\href@noop {} {\bibfield  {journal} {\bibinfo  {journal} {Applied Physics Reviews}\ }\textbf {\bibinfo {volume} {9}},\ \bibinfo {pages} {011411} (\bibinfo {year} {2022})}\BibitemShut {NoStop}%
\bibitem [{\citenamefont {Yang}\ \emph {et~al.}(2020)\citenamefont {Yang}, \citenamefont {Wang}, \citenamefont {Liu}, \citenamefont {Xu}, \citenamefont {Li}, \citenamefont {Xia}, \citenamefont {Li},\ and\ \citenamefont {Gao}}]{yang20}%
  \BibitemOpen
  \bibfield  {author} {\bibinfo {author} {\bibfnamefont {J.}~\bibnamefont {Yang}}, \bibinfo {author} {\bibfnamefont {J.}~\bibnamefont {Wang}}, \bibinfo {author} {\bibfnamefont {Q.}~\bibnamefont {Liu}}, \bibinfo {author} {\bibfnamefont {R.}~\bibnamefont {Xu}}, \bibinfo {author} {\bibfnamefont {Y.}~\bibnamefont {Li}}, \bibinfo {author} {\bibfnamefont {M.}~\bibnamefont {Xia}}, \bibinfo {author} {\bibfnamefont {Z.}~\bibnamefont {Li}},\ and\ \bibinfo {author} {\bibfnamefont {F.}~\bibnamefont {Gao}},\ }\bibfield  {title} {\bibinfo {title} {Enhancement of ferromagnetism for vi3 monolayer},\ }\href@noop {} {\bibfield  {journal} {\bibinfo  {journal} {Applied Surface Science}\ }\textbf {\bibinfo {volume} {524}},\ \bibinfo {pages} {146490} (\bibinfo {year} {2020})}\BibitemShut {NoStop}%
\bibitem [{\citenamefont {Jahn}\ and\ \citenamefont {Teller}(1937)}]{jahn37}%
  \BibitemOpen
  \bibfield  {author} {\bibinfo {author} {\bibfnamefont {H.~A.}\ \bibnamefont {Jahn}}\ and\ \bibinfo {author} {\bibfnamefont {E.}~\bibnamefont {Teller}},\ }\bibfield  {title} {\bibinfo {title} {Stability of polyatomic molecules in degenerate electronic states-i—orbital degeneracy},\ }\href@noop {} {\bibfield  {journal} {\bibinfo  {journal} {Proceedings of the Royal Society of London. Series A-Mathematical and Physical Sciences}\ }\textbf {\bibinfo {volume} {161}},\ \bibinfo {pages} {220} (\bibinfo {year} {1937})}\BibitemShut {NoStop}%
\bibitem [{\citenamefont {Zhang}\ \emph {et~al.}(2020{\natexlab{a}})\citenamefont {Zhang}, \citenamefont {Yang}, \citenamefont {Cui}, \citenamefont {Xu},\ and\ \citenamefont {Zhang}}]{zhang20-CrMnI6}%
  \BibitemOpen
  \bibfield  {author} {\bibinfo {author} {\bibfnamefont {H.}~\bibnamefont {Zhang}}, \bibinfo {author} {\bibfnamefont {W.}~\bibnamefont {Yang}}, \bibinfo {author} {\bibfnamefont {P.}~\bibnamefont {Cui}}, \bibinfo {author} {\bibfnamefont {X.}~\bibnamefont {Xu}},\ and\ \bibinfo {author} {\bibfnamefont {Z.}~\bibnamefont {Zhang}},\ }\bibfield  {title} {\bibinfo {title} {Prediction of monolayered ferromagnetic crmni 6 as an intrinsic high-temperature quantum anomalous hall system},\ }\href@noop {} {\bibfield  {journal} {\bibinfo  {journal} {Physical Review B}\ }\textbf {\bibinfo {volume} {102}},\ \bibinfo {pages} {115413} (\bibinfo {year} {2020}{\natexlab{a}})}\BibitemShut {NoStop}%
\bibitem [{\citenamefont {Li}\ \emph {et~al.}(2024)\citenamefont {Li}, \citenamefont {Liu}, \citenamefont {Zhang}, \citenamefont {Zhang}, \citenamefont {Zhang}, \citenamefont {Zhang}, \citenamefont {Meng}, \citenamefont {Hou}, \citenamefont {Li}, \citenamefont {Kang} \emph {et~al.}}]{li24}%
  \BibitemOpen
  \bibfield  {author} {\bibinfo {author} {\bibfnamefont {X.}~\bibnamefont {Li}}, \bibinfo {author} {\bibfnamefont {C.}~\bibnamefont {Liu}}, \bibinfo {author} {\bibfnamefont {Y.}~\bibnamefont {Zhang}}, \bibinfo {author} {\bibfnamefont {S.}~\bibnamefont {Zhang}}, \bibinfo {author} {\bibfnamefont {H.}~\bibnamefont {Zhang}}, \bibinfo {author} {\bibfnamefont {Y.}~\bibnamefont {Zhang}}, \bibinfo {author} {\bibfnamefont {W.}~\bibnamefont {Meng}}, \bibinfo {author} {\bibfnamefont {D.}~\bibnamefont {Hou}}, \bibinfo {author} {\bibfnamefont {T.}~\bibnamefont {Li}}, \bibinfo {author} {\bibfnamefont {C.}~\bibnamefont {Kang}}, \emph {et~al.},\ }\bibfield  {title} {\bibinfo {title} {Topological kerr effects in two-dimensional magnets with broken inversion symmetry},\ }\href@noop {} {\bibfield  {journal} {\bibinfo  {journal} {Nature Physics}\ }\textbf {\bibinfo {volume} {20}},\ \bibinfo {pages} {1145} (\bibinfo {year} {2024})}\BibitemShut {NoStop}%
\bibitem [{\citenamefont {{\v{Z}}uti{\'c}}\ \emph {et~al.}(2004)\citenamefont {{\v{Z}}uti{\'c}}, \citenamefont {Fabian},\ and\ \citenamefont {Sarma}}]{Igor04}%
  \BibitemOpen
  \bibfield  {author} {\bibinfo {author} {\bibfnamefont {I.}~\bibnamefont {{\v{Z}}uti{\'c}}}, \bibinfo {author} {\bibfnamefont {J.}~\bibnamefont {Fabian}},\ and\ \bibinfo {author} {\bibfnamefont {S.~D.}\ \bibnamefont {Sarma}},\ }\bibfield  {title} {\bibinfo {title} {Spintronics: Fundamentals and applications},\ }\href@noop {} {\bibfield  {journal} {\bibinfo  {journal} {Reviews of modern physics}\ }\textbf {\bibinfo {volume} {76}},\ \bibinfo {pages} {323} (\bibinfo {year} {2004})}\BibitemShut {NoStop}%
\bibitem [{\citenamefont {Schmidt}\ \emph {et~al.}(2000)\citenamefont {Schmidt}, \citenamefont {Ferrand}, \citenamefont {Molenkamp}, \citenamefont {Filip},\ and\ \citenamefont {Van~Wees}}]{SFET-Schmidt}%
  \BibitemOpen
  \bibfield  {author} {\bibinfo {author} {\bibfnamefont {G.}~\bibnamefont {Schmidt}}, \bibinfo {author} {\bibfnamefont {D.}~\bibnamefont {Ferrand}}, \bibinfo {author} {\bibfnamefont {L.}~\bibnamefont {Molenkamp}}, \bibinfo {author} {\bibfnamefont {A.}~\bibnamefont {Filip}},\ and\ \bibinfo {author} {\bibfnamefont {B.}~\bibnamefont {Van~Wees}},\ }\bibfield  {title} {\bibinfo {title} {Fundamental obstacle for electrical spin injection from a ferromagnetic metal into a diffusive semiconductor},\ }\href@noop {} {\bibfield  {journal} {\bibinfo  {journal} {Physical Review B}\ }\textbf {\bibinfo {volume} {62}},\ \bibinfo {pages} {R4790} (\bibinfo {year} {2000})}\BibitemShut {NoStop}%
\bibitem [{\citenamefont {Xu}\ \emph {et~al.}(2015)\citenamefont {Xu}, \citenamefont {Zhang}, \citenamefont {Hou}, \citenamefont {Wang}, \citenamefont {Liu}, \citenamefont {Xi}, \citenamefont {Wang}, \citenamefont {Wang}, \citenamefont {Luo}, \citenamefont {Wang} \emph {et~al.}}]{Xu15}%
  \BibitemOpen
  \bibfield  {author} {\bibinfo {author} {\bibfnamefont {G.}~\bibnamefont {Xu}}, \bibinfo {author} {\bibfnamefont {X.}~\bibnamefont {Zhang}}, \bibinfo {author} {\bibfnamefont {Z.}~\bibnamefont {Hou}}, \bibinfo {author} {\bibfnamefont {Y.}~\bibnamefont {Wang}}, \bibinfo {author} {\bibfnamefont {E.}~\bibnamefont {Liu}}, \bibinfo {author} {\bibfnamefont {X.}~\bibnamefont {Xi}}, \bibinfo {author} {\bibfnamefont {S.}~\bibnamefont {Wang}}, \bibinfo {author} {\bibfnamefont {W.}~\bibnamefont {Wang}}, \bibinfo {author} {\bibfnamefont {H.}~\bibnamefont {Luo}}, \bibinfo {author} {\bibfnamefont {W.}~\bibnamefont {Wang}}, \emph {et~al.},\ }\bibfield  {title} {\bibinfo {title} {New spin injection scheme based on spin gapless semiconductors: A first-principles study},\ }\href@noop {} {\bibfield  {journal} {\bibinfo  {journal} {Europhysics Letters}\ }\textbf {\bibinfo {volume} {111}},\ \bibinfo {pages} {68003} (\bibinfo {year} {2015})}\BibitemShut {NoStop}%
\bibitem [{\citenamefont {Ezawa}(2013)}]{ezawa13}%
  \BibitemOpen
  \bibfield  {author} {\bibinfo {author} {\bibfnamefont {M.}~\bibnamefont {Ezawa}},\ }\bibfield  {title} {\bibinfo {title} {Quantized conductance and field-effect topological quantum transistor in silicene nanoribbons},\ }\href@noop {} {\bibfield  {journal} {\bibinfo  {journal} {Applied Physics Letters}\ }\textbf {\bibinfo {volume} {102}},\ \bibinfo {pages} {172103} (\bibinfo {year} {2013})}\BibitemShut {NoStop}%
\bibitem [{\citenamefont {Zhang}\ \emph {et~al.}(2017)\citenamefont {Zhang}, \citenamefont {Feng}, \citenamefont {Wang}, \citenamefont {Lian}, \citenamefont {Zhang}, \citenamefont {Chang}, \citenamefont {Guo}, \citenamefont {Ou}, \citenamefont {Feng}, \citenamefont {Zhang}, \citenamefont {He}, \citenamefont {Ma}, \citenamefont {Xue},\ and\ \citenamefont {Wang}}]{zhang17}%
  \BibitemOpen
  \bibfield  {author} {\bibinfo {author} {\bibfnamefont {Z.}~\bibnamefont {Zhang}}, \bibinfo {author} {\bibfnamefont {X.}~\bibnamefont {Feng}}, \bibinfo {author} {\bibfnamefont {J.}~\bibnamefont {Wang}}, \bibinfo {author} {\bibfnamefont {B.}~\bibnamefont {Lian}}, \bibinfo {author} {\bibfnamefont {J.}~\bibnamefont {Zhang}}, \bibinfo {author} {\bibfnamefont {C.}~\bibnamefont {Chang}}, \bibinfo {author} {\bibfnamefont {M.}~\bibnamefont {Guo}}, \bibinfo {author} {\bibfnamefont {Y.}~\bibnamefont {Ou}}, \bibinfo {author} {\bibfnamefont {Y.}~\bibnamefont {Feng}}, \bibinfo {author} {\bibfnamefont {S.-C.}\ \bibnamefont {Zhang}}, \bibinfo {author} {\bibfnamefont {K.}~\bibnamefont {He}}, \bibinfo {author} {\bibfnamefont {X.}~\bibnamefont {Ma}}, \bibinfo {author} {\bibfnamefont {Q.-K.}\ \bibnamefont {Xue}},\ and\ \bibinfo {author} {\bibfnamefont {Y.}~\bibnamefont {Wang}},\ }\bibfield  {title} {\bibinfo {title} {Magnetic quantum phase transition in cr-doped $bi_{2}(se_{x} te_{1-x})_{3}$ driven by the stark effect},\
  }\href@noop {} {\bibfield  {journal} {\bibinfo  {journal} {Nature nanotechnology}\ }\textbf {\bibinfo {volume} {12}},\ \bibinfo {pages} {953} (\bibinfo {year} {2017})}\BibitemShut {NoStop}%
\bibitem [{\citenamefont {Xu}\ \emph {et~al.}(2019)\citenamefont {Xu}, \citenamefont {Chen}, \citenamefont {Wang}, \citenamefont {Liu},\ and\ \citenamefont {Ma}}]{xu19-TTFET}%
  \BibitemOpen
  \bibfield  {author} {\bibinfo {author} {\bibfnamefont {Y.}~\bibnamefont {Xu}}, \bibinfo {author} {\bibfnamefont {Y.-R.}\ \bibnamefont {Chen}}, \bibinfo {author} {\bibfnamefont {J.}~\bibnamefont {Wang}}, \bibinfo {author} {\bibfnamefont {J.-F.}\ \bibnamefont {Liu}},\ and\ \bibinfo {author} {\bibfnamefont {Z.}~\bibnamefont {Ma}},\ }\bibfield  {title} {\bibinfo {title} {Quantized field-effect tunneling between topological edge or interface states},\ }\href@noop {} {\bibfield  {journal} {\bibinfo  {journal} {Physical Review Letters}\ }\textbf {\bibinfo {volume} {123}},\ \bibinfo {pages} {206801} (\bibinfo {year} {2019})}\BibitemShut {NoStop}%
\bibitem [{\citenamefont {Nadeem}\ \emph {et~al.}(2021)\citenamefont {Nadeem}, \citenamefont {Di~Bernardo}, \citenamefont {Wang}, \citenamefont {Fuhrer},\ and\ \citenamefont {Culcer}}]{nadeem21}%
  \BibitemOpen
  \bibfield  {author} {\bibinfo {author} {\bibfnamefont {M.}~\bibnamefont {Nadeem}}, \bibinfo {author} {\bibfnamefont {I.}~\bibnamefont {Di~Bernardo}}, \bibinfo {author} {\bibfnamefont {X.}~\bibnamefont {Wang}}, \bibinfo {author} {\bibfnamefont {M.~S.}\ \bibnamefont {Fuhrer}},\ and\ \bibinfo {author} {\bibfnamefont {D.}~\bibnamefont {Culcer}},\ }\bibfield  {title} {\bibinfo {title} {Overcoming boltzmann’s tyranny in a transistor via the topological quantum field effect},\ }\href@noop {} {\bibfield  {journal} {\bibinfo  {journal} {Nano Letters}\ }\textbf {\bibinfo {volume} {21}},\ \bibinfo {pages} {3155} (\bibinfo {year} {2021})}\BibitemShut {NoStop}%
\bibitem [{\citenamefont {Fuhrer}\ \emph {et~al.}(2021)\citenamefont {Fuhrer}, \citenamefont {Edmonds}, \citenamefont {Culcer}, \citenamefont {Nadeem}, \citenamefont {Wang}, \citenamefont {Medhekar}, \citenamefont {Yin},\ and\ \citenamefont {Cole}}]{fuhrer21}%
  \BibitemOpen
  \bibfield  {author} {\bibinfo {author} {\bibfnamefont {M.~S.}\ \bibnamefont {Fuhrer}}, \bibinfo {author} {\bibfnamefont {M.~T.}\ \bibnamefont {Edmonds}}, \bibinfo {author} {\bibfnamefont {D.}~\bibnamefont {Culcer}}, \bibinfo {author} {\bibfnamefont {M.}~\bibnamefont {Nadeem}}, \bibinfo {author} {\bibfnamefont {X.}~\bibnamefont {Wang}}, \bibinfo {author} {\bibfnamefont {N.}~\bibnamefont {Medhekar}}, \bibinfo {author} {\bibfnamefont {Y.}~\bibnamefont {Yin}},\ and\ \bibinfo {author} {\bibfnamefont {J.~H.}\ \bibnamefont {Cole}},\ }\bibfield  {title} {\bibinfo {title} {Proposal for a negative capacitance topological quantum field-effect transistor},\ }in\ \href@noop {} {\emph {\bibinfo {booktitle} {2021 IEEE International Electron Devices Meeting (IEDM)}}}\ (\bibinfo {organization} {IEEE},\ \bibinfo {year} {2021})\ pp.\ \bibinfo {pages} {38--2}\BibitemShut {NoStop}%
\bibitem [{\citenamefont {Banerjee}\ \emph {et~al.}(2022)\citenamefont {Banerjee}, \citenamefont {Jana}, \citenamefont {Basak}, \citenamefont {Fuhrer}, \citenamefont {Culcer},\ and\ \citenamefont {Muralidharan}}]{banerjee22}%
  \BibitemOpen
  \bibfield  {author} {\bibinfo {author} {\bibfnamefont {S.}~\bibnamefont {Banerjee}}, \bibinfo {author} {\bibfnamefont {K.}~\bibnamefont {Jana}}, \bibinfo {author} {\bibfnamefont {A.}~\bibnamefont {Basak}}, \bibinfo {author} {\bibfnamefont {M.~S.}\ \bibnamefont {Fuhrer}}, \bibinfo {author} {\bibfnamefont {D.}~\bibnamefont {Culcer}},\ and\ \bibinfo {author} {\bibfnamefont {B.}~\bibnamefont {Muralidharan}},\ }\bibfield  {title} {\bibinfo {title} {Robust subthermionic topological transistor action via antiferromagnetic exchange},\ }\href@noop {} {\bibfield  {journal} {\bibinfo  {journal} {Physical Review Applied}\ }\textbf {\bibinfo {volume} {18}},\ \bibinfo {pages} {054088} (\bibinfo {year} {2022})}\BibitemShut {NoStop}%
\bibitem [{\citenamefont {Weber}\ \emph {et~al.}(2024)\citenamefont {Weber}, \citenamefont {Fuhrer}, \citenamefont {Sheng}, \citenamefont {Yang}, \citenamefont {Thomale}, \citenamefont {Shamim}, \citenamefont {Molenkamp}, \citenamefont {Cobden}, \citenamefont {Pesin}, \citenamefont {Zandvliet} \emph {et~al.}}]{weber24}%
  \BibitemOpen
  \bibfield  {author} {\bibinfo {author} {\bibfnamefont {B.}~\bibnamefont {Weber}}, \bibinfo {author} {\bibfnamefont {M.~S.}\ \bibnamefont {Fuhrer}}, \bibinfo {author} {\bibfnamefont {X.-L.}\ \bibnamefont {Sheng}}, \bibinfo {author} {\bibfnamefont {S.~A.}\ \bibnamefont {Yang}}, \bibinfo {author} {\bibfnamefont {R.}~\bibnamefont {Thomale}}, \bibinfo {author} {\bibfnamefont {S.}~\bibnamefont {Shamim}}, \bibinfo {author} {\bibfnamefont {L.~W.}\ \bibnamefont {Molenkamp}}, \bibinfo {author} {\bibfnamefont {D.}~\bibnamefont {Cobden}}, \bibinfo {author} {\bibfnamefont {D.}~\bibnamefont {Pesin}}, \bibinfo {author} {\bibfnamefont {H.~J.}\ \bibnamefont {Zandvliet}}, \emph {et~al.},\ }\bibfield  {title} {\bibinfo {title} {2024 roadmap on 2d topological insulators},\ }\href@noop {} {\bibfield  {journal} {\bibinfo  {journal} {Journal of Physics: Materials}\ }\textbf {\bibinfo {volume} {7}},\ \bibinfo {pages} {022501} (\bibinfo {year} {2024})}\BibitemShut {NoStop}%
\bibitem [{\citenamefont {Tran}\ and\ \citenamefont {Matsushita}(2024)}]{tran24}%
  \BibitemOpen
  \bibfield  {author} {\bibinfo {author} {\bibfnamefont {H.~B.}\ \bibnamefont {Tran}}\ and\ \bibinfo {author} {\bibfnamefont {Y.-i.}\ \bibnamefont {Matsushita}},\ }\bibfield  {title} {\bibinfo {title} {Skyrmions in van der waals centrosymmetric materials with dzyaloshinskii--moriya interactions},\ }\href@noop {} {\bibfield  {journal} {\bibinfo  {journal} {Scripta Materialia}\ }\textbf {\bibinfo {volume} {239}},\ \bibinfo {pages} {115799} (\bibinfo {year} {2024})}\BibitemShut {NoStop}%
\bibitem [{\citenamefont {Pan}\ \emph {et~al.}(2014)\citenamefont {Pan}, \citenamefont {Li}, \citenamefont {Liu}, \citenamefont {Zhu}, \citenamefont {Qiao},\ and\ \citenamefont {Yao}}]{pan14}%
  \BibitemOpen
  \bibfield  {author} {\bibinfo {author} {\bibfnamefont {H.}~\bibnamefont {Pan}}, \bibinfo {author} {\bibfnamefont {Z.}~\bibnamefont {Li}}, \bibinfo {author} {\bibfnamefont {C.-C.}\ \bibnamefont {Liu}}, \bibinfo {author} {\bibfnamefont {G.}~\bibnamefont {Zhu}}, \bibinfo {author} {\bibfnamefont {Z.}~\bibnamefont {Qiao}},\ and\ \bibinfo {author} {\bibfnamefont {Y.}~\bibnamefont {Yao}},\ }\bibfield  {title} {\bibinfo {title} {Valley-polarized quantum anomalous hall effect in silicene},\ }\href@noop {} {\bibfield  {journal} {\bibinfo  {journal} {Physical review letters}\ }\textbf {\bibinfo {volume} {112}},\ \bibinfo {pages} {106802} (\bibinfo {year} {2014})}\BibitemShut {NoStop}%
\bibitem [{\citenamefont {Zhou}\ \emph {et~al.}(2017)\citenamefont {Zhou}, \citenamefont {Sun},\ and\ \citenamefont {Jena}}]{zhou17}%
  \BibitemOpen
  \bibfield  {author} {\bibinfo {author} {\bibfnamefont {J.}~\bibnamefont {Zhou}}, \bibinfo {author} {\bibfnamefont {Q.}~\bibnamefont {Sun}},\ and\ \bibinfo {author} {\bibfnamefont {P.}~\bibnamefont {Jena}},\ }\bibfield  {title} {\bibinfo {title} {Valley-polarized quantum anomalous hall effect in ferrimagnetic honeycomb lattices},\ }\href@noop {} {\bibfield  {journal} {\bibinfo  {journal} {Physical Review Letters}\ }\textbf {\bibinfo {volume} {119}},\ \bibinfo {pages} {046403} (\bibinfo {year} {2017})}\BibitemShut {NoStop}%
\bibitem [{\citenamefont {Zhang}\ \emph {et~al.}(2020{\natexlab{b}})\citenamefont {Zhang}, \citenamefont {Yang}, \citenamefont {Ning},\ and\ \citenamefont {Xu}}]{zhang20}%
  \BibitemOpen
  \bibfield  {author} {\bibinfo {author} {\bibfnamefont {H.}~\bibnamefont {Zhang}}, \bibinfo {author} {\bibfnamefont {W.}~\bibnamefont {Yang}}, \bibinfo {author} {\bibfnamefont {Y.}~\bibnamefont {Ning}},\ and\ \bibinfo {author} {\bibfnamefont {X.}~\bibnamefont {Xu}},\ }\bibfield  {title} {\bibinfo {title} {Abundant valley-polarized states in two-dimensional ferromagnetic van der waals heterostructures},\ }\href@noop {} {\bibfield  {journal} {\bibinfo  {journal} {Physical Review B}\ }\textbf {\bibinfo {volume} {101}},\ \bibinfo {pages} {205404} (\bibinfo {year} {2020}{\natexlab{b}})}\BibitemShut {NoStop}%
\bibitem [{\citenamefont {Huang}\ \emph {et~al.}(2017)\citenamefont {Huang}, \citenamefont {Zhou}, \citenamefont {Wu}, \citenamefont {Deng}, \citenamefont {Jena},\ and\ \citenamefont {Kan}}]{RuI}%
  \BibitemOpen
  \bibfield  {author} {\bibinfo {author} {\bibfnamefont {C.}~\bibnamefont {Huang}}, \bibinfo {author} {\bibfnamefont {J.}~\bibnamefont {Zhou}}, \bibinfo {author} {\bibfnamefont {H.}~\bibnamefont {Wu}}, \bibinfo {author} {\bibfnamefont {K.}~\bibnamefont {Deng}}, \bibinfo {author} {\bibfnamefont {P.}~\bibnamefont {Jena}},\ and\ \bibinfo {author} {\bibfnamefont {E.}~\bibnamefont {Kan}},\ }\bibfield  {title} {\bibinfo {title} {Quantum anomalous hall effect in ferromagnetic transition metal halides},\ }\href@noop {} {\bibfield  {journal} {\bibinfo  {journal} {Physical Review B}\ }\textbf {\bibinfo {volume} {95}},\ \bibinfo {pages} {045113} (\bibinfo {year} {2017})}\BibitemShut {NoStop}%
\bibitem [{\citenamefont {Sun}\ and\ \citenamefont {Kioussis}(2018)}]{MnX}%
  \BibitemOpen
  \bibfield  {author} {\bibinfo {author} {\bibfnamefont {Q.}~\bibnamefont {Sun}}\ and\ \bibinfo {author} {\bibfnamefont {N.}~\bibnamefont {Kioussis}},\ }\bibfield  {title} {\bibinfo {title} {Prediction of manganese trihalides as two-dimensional dirac half-metals},\ }\href@noop {} {\bibfield  {journal} {\bibinfo  {journal} {Physical Review B}\ }\textbf {\bibinfo {volume} {97}},\ \bibinfo {pages} {094408} (\bibinfo {year} {2018})}\BibitemShut {NoStop}%
\bibitem [{\citenamefont {He}\ \emph {et~al.}(2017)\citenamefont {He}, \citenamefont {Li}, \citenamefont {Lyu},\ and\ \citenamefont {Nachtigall}}]{NiCl}%
  \BibitemOpen
  \bibfield  {author} {\bibinfo {author} {\bibfnamefont {J.}~\bibnamefont {He}}, \bibinfo {author} {\bibfnamefont {X.}~\bibnamefont {Li}}, \bibinfo {author} {\bibfnamefont {P.}~\bibnamefont {Lyu}},\ and\ \bibinfo {author} {\bibfnamefont {P.}~\bibnamefont {Nachtigall}},\ }\bibfield  {title} {\bibinfo {title} {Near-room-temperature chern insulator and dirac spin-gapless semiconductor: nickel chloride monolayer},\ }\href@noop {} {\bibfield  {journal} {\bibinfo  {journal} {Nanoscale}\ }\textbf {\bibinfo {volume} {9}},\ \bibinfo {pages} {2246} (\bibinfo {year} {2017})}\BibitemShut {NoStop}%
\bibitem [{\citenamefont {You}\ \emph {et~al.}(2019{\natexlab{a}})\citenamefont {You}, \citenamefont {Chen}, \citenamefont {Zhang}, \citenamefont {Sheng}, \citenamefont {Yang},\ and\ \citenamefont {Su}}]{PtCl-You19}%
  \BibitemOpen
  \bibfield  {author} {\bibinfo {author} {\bibfnamefont {J.-Y.}\ \bibnamefont {You}}, \bibinfo {author} {\bibfnamefont {C.}~\bibnamefont {Chen}}, \bibinfo {author} {\bibfnamefont {Z.}~\bibnamefont {Zhang}}, \bibinfo {author} {\bibfnamefont {X.-L.}\ \bibnamefont {Sheng}}, \bibinfo {author} {\bibfnamefont {S.~A.}\ \bibnamefont {Yang}},\ and\ \bibinfo {author} {\bibfnamefont {G.}~\bibnamefont {Su}},\ }\bibfield  {title} {\bibinfo {title} {Two-dimensional weyl half-semimetal and tunable quantum anomalous hall effect},\ }\href@noop {} {\bibfield  {journal} {\bibinfo  {journal} {Physical Review B}\ }\textbf {\bibinfo {volume} {100}},\ \bibinfo {pages} {064408} (\bibinfo {year} {2019}{\natexlab{a}})}\BibitemShut {NoStop}%
\bibitem [{\citenamefont {Wang}\ \emph {et~al.}(2018)\citenamefont {Wang}, \citenamefont {Li}, \citenamefont {Zhang}, \citenamefont {Zhang}, \citenamefont {Ji}, \citenamefont {Li},\ and\ \citenamefont {Wang}}]{wang18-pdcl}%
  \BibitemOpen
  \bibfield  {author} {\bibinfo {author} {\bibfnamefont {Y.-p.}\ \bibnamefont {Wang}}, \bibinfo {author} {\bibfnamefont {S.-s.}\ \bibnamefont {Li}}, \bibinfo {author} {\bibfnamefont {C.-w.}\ \bibnamefont {Zhang}}, \bibinfo {author} {\bibfnamefont {S.-f.}\ \bibnamefont {Zhang}}, \bibinfo {author} {\bibfnamefont {W.-x.}\ \bibnamefont {Ji}}, \bibinfo {author} {\bibfnamefont {P.}~\bibnamefont {Li}},\ and\ \bibinfo {author} {\bibfnamefont {P.-j.}\ \bibnamefont {Wang}},\ }\bibfield  {title} {\bibinfo {title} {High-temperature dirac half-metal pdcl 3: a promising candidate for realizing quantum anomalous hall effect},\ }\href@noop {} {\bibfield  {journal} {\bibinfo  {journal} {Journal of Materials Chemistry C}\ }\textbf {\bibinfo {volume} {6}},\ \bibinfo {pages} {10284} (\bibinfo {year} {2018})}\BibitemShut {NoStop}%
\bibitem [{\citenamefont {You}\ \emph {et~al.}(2019{\natexlab{b}})\citenamefont {You}, \citenamefont {Zhang}, \citenamefont {Gu},\ and\ \citenamefont {Su}}]{you19-PRA}%
  \BibitemOpen
  \bibfield  {author} {\bibinfo {author} {\bibfnamefont {J.-Y.}\ \bibnamefont {You}}, \bibinfo {author} {\bibfnamefont {Z.}~\bibnamefont {Zhang}}, \bibinfo {author} {\bibfnamefont {B.}~\bibnamefont {Gu}},\ and\ \bibinfo {author} {\bibfnamefont {G.}~\bibnamefont {Su}},\ }\bibfield  {title} {\bibinfo {title} {Two-dimensional room-temperature ferromagnetic semiconductors with quantum anomalous hall effect},\ }\href@noop {} {\bibfield  {journal} {\bibinfo  {journal} {Physical Review Applied}\ }\textbf {\bibinfo {volume} {12}},\ \bibinfo {pages} {024063} (\bibinfo {year} {2019}{\natexlab{b}})}\BibitemShut {NoStop}%
\bibitem [{\citenamefont {Sun}\ \emph {et~al.}(2020)\citenamefont {Sun}, \citenamefont {Zhong}, \citenamefont {Cui}, \citenamefont {Shi}, \citenamefont {Hao}, \citenamefont {Xu},\ and\ \citenamefont {Li}}]{sun20}%
  \BibitemOpen
  \bibfield  {author} {\bibinfo {author} {\bibfnamefont {J.}~\bibnamefont {Sun}}, \bibinfo {author} {\bibfnamefont {X.}~\bibnamefont {Zhong}}, \bibinfo {author} {\bibfnamefont {W.}~\bibnamefont {Cui}}, \bibinfo {author} {\bibfnamefont {J.}~\bibnamefont {Shi}}, \bibinfo {author} {\bibfnamefont {J.}~\bibnamefont {Hao}}, \bibinfo {author} {\bibfnamefont {M.}~\bibnamefont {Xu}},\ and\ \bibinfo {author} {\bibfnamefont {Y.}~\bibnamefont {Li}},\ }\bibfield  {title} {\bibinfo {title} {The intrinsic magnetism, quantum anomalous hall effect and curie temperature in 2d transition metal trihalides},\ }\href@noop {} {\bibfield  {journal} {\bibinfo  {journal} {Physical Chemistry Chemical Physics}\ }\textbf {\bibinfo {volume} {22}},\ \bibinfo {pages} {2429} (\bibinfo {year} {2020})}\BibitemShut {NoStop}%
\bibitem [{\citenamefont {Li}(2019)}]{li19}%
  \BibitemOpen
  \bibfield  {author} {\bibinfo {author} {\bibfnamefont {P.}~\bibnamefont {Li}},\ }\bibfield  {title} {\bibinfo {title} {Prediction of intrinsic two dimensional ferromagnetism realized quantum anomalous hall effect},\ }\href@noop {} {\bibfield  {journal} {\bibinfo  {journal} {Physical Chemistry Chemical Physics}\ }\textbf {\bibinfo {volume} {21}},\ \bibinfo {pages} {6712} (\bibinfo {year} {2019})}\BibitemShut {NoStop}%
\bibitem [{\citenamefont {Jiang}\ \emph {et~al.}(2021)\citenamefont {Jiang}, \citenamefont {Yu}, \citenamefont {Cui}, \citenamefont {Liu}, \citenamefont {Xie}, \citenamefont {Liao}, \citenamefont {Zhang}, \citenamefont {Huang}, \citenamefont {Ning}, \citenamefont {Jia} \emph {et~al.}}]{jiang21}%
  \BibitemOpen
  \bibfield  {author} {\bibinfo {author} {\bibfnamefont {B.}~\bibnamefont {Jiang}}, \bibinfo {author} {\bibfnamefont {Y.}~\bibnamefont {Yu}}, \bibinfo {author} {\bibfnamefont {J.}~\bibnamefont {Cui}}, \bibinfo {author} {\bibfnamefont {X.}~\bibnamefont {Liu}}, \bibinfo {author} {\bibfnamefont {L.}~\bibnamefont {Xie}}, \bibinfo {author} {\bibfnamefont {J.}~\bibnamefont {Liao}}, \bibinfo {author} {\bibfnamefont {Q.}~\bibnamefont {Zhang}}, \bibinfo {author} {\bibfnamefont {Y.}~\bibnamefont {Huang}}, \bibinfo {author} {\bibfnamefont {S.}~\bibnamefont {Ning}}, \bibinfo {author} {\bibfnamefont {B.}~\bibnamefont {Jia}}, \emph {et~al.},\ }\bibfield  {title} {\bibinfo {title} {High-entropy-stabilized chalcogenides with high thermoelectric performance},\ }\href@noop {} {\bibfield  {journal} {\bibinfo  {journal} {Science}\ }\textbf {\bibinfo {volume} {371}},\ \bibinfo {pages} {830} (\bibinfo {year} {2021})}\BibitemShut {NoStop}%
\bibitem [{\citenamefont {B.~Bychkov}\ and\ \citenamefont {Rasbha}(1984)}]{Bychkov-SOI84}%
  \BibitemOpen
  \bibfield  {author} {\bibinfo {author} {\bibfnamefont {Y.~A.}\ \bibnamefont {B.~Bychkov}}\ and\ \bibinfo {author} {\bibfnamefont {E.~I.}\ \bibnamefont {Rasbha}},\ }\bibfield  {title} {\bibinfo {title} {Properties of a 2d electron gas with lifted spectral degeneracy},\ }\href@noop {} {\bibfield  {journal} {\bibinfo  {journal} {P. Zh. Eksp. Teor. Fiz.}\ }\textbf {\bibinfo {volume} {39}},\ \bibinfo {pages} {66–69} (\bibinfo {year} {1984})}\BibitemShut {NoStop}%
\bibitem [{\citenamefont {Manchon}\ \emph {et~al.}(2015)\citenamefont {Manchon}, \citenamefont {Koo}, \citenamefont {Nitta}, \citenamefont {Frolov},\ and\ \citenamefont {Duine}}]{manchon15}%
  \BibitemOpen
  \bibfield  {author} {\bibinfo {author} {\bibfnamefont {A.}~\bibnamefont {Manchon}}, \bibinfo {author} {\bibfnamefont {H.~C.}\ \bibnamefont {Koo}}, \bibinfo {author} {\bibfnamefont {J.}~\bibnamefont {Nitta}}, \bibinfo {author} {\bibfnamefont {S.~M.}\ \bibnamefont {Frolov}},\ and\ \bibinfo {author} {\bibfnamefont {R.~A.}\ \bibnamefont {Duine}},\ }\bibfield  {title} {\bibinfo {title} {New perspectives for rashba spin--orbit coupling},\ }\href@noop {} {\bibfield  {journal} {\bibinfo  {journal} {Nature materials}\ }\textbf {\bibinfo {volume} {14}},\ \bibinfo {pages} {871} (\bibinfo {year} {2015})}\BibitemShut {NoStop}%
\bibitem [{\citenamefont {Anandan}\ \emph {et~al.}(2023)\citenamefont {Anandan}, \citenamefont {Nadeem}, \citenamefont {Lin}, \citenamefont {Singh}, \citenamefont {Guan}, \citenamefont {Kim}, \citenamefont {Shahrokhi}, \citenamefont {Rahaman}, \citenamefont {Geng}, \citenamefont {Huang} \emph {et~al.}}]{anandan23}%
  \BibitemOpen
  \bibfield  {author} {\bibinfo {author} {\bibfnamefont {P.~R.}\ \bibnamefont {Anandan}}, \bibinfo {author} {\bibfnamefont {M.}~\bibnamefont {Nadeem}}, \bibinfo {author} {\bibfnamefont {C.-H.}\ \bibnamefont {Lin}}, \bibinfo {author} {\bibfnamefont {S.}~\bibnamefont {Singh}}, \bibinfo {author} {\bibfnamefont {X.}~\bibnamefont {Guan}}, \bibinfo {author} {\bibfnamefont {J.}~\bibnamefont {Kim}}, \bibinfo {author} {\bibfnamefont {S.}~\bibnamefont {Shahrokhi}}, \bibinfo {author} {\bibfnamefont {M.~Z.}\ \bibnamefont {Rahaman}}, \bibinfo {author} {\bibfnamefont {X.}~\bibnamefont {Geng}}, \bibinfo {author} {\bibfnamefont {J.-K.}\ \bibnamefont {Huang}}, \emph {et~al.},\ }\bibfield  {title} {\bibinfo {title} {Spin--orbital coupling in all-inorganic metal-halide perovskites: The hidden force that matters},\ }\href@noop {} {\bibfield  {journal} {\bibinfo  {journal} {Applied Physics Reviews}\ }\textbf {\bibinfo {volume} {10}} (\bibinfo {year} {2023})}\BibitemShut {NoStop}%
\bibitem [{\citenamefont {Rashba}(2009)}]{rashba09}%
  \BibitemOpen
  \bibfield  {author} {\bibinfo {author} {\bibfnamefont {E.~I.}\ \bibnamefont {Rashba}},\ }\bibfield  {title} {\bibinfo {title} {Graphene with structure-induced spin-orbit coupling: Spin-polarized states, spin zero modes, and quantum hall effect},\ }\href@noop {} {\bibfield  {journal} {\bibinfo  {journal} {Physical Review B}\ }\textbf {\bibinfo {volume} {79}},\ \bibinfo {pages} {161409} (\bibinfo {year} {2009})}\BibitemShut {NoStop}%
\bibitem [{\citenamefont {Nadeem}\ \emph {et~al.}(2023{\natexlab{a}})\citenamefont {Nadeem}, \citenamefont {Fuhrer},\ and\ \citenamefont {Wang}}]{nadeem23-NP}%
  \BibitemOpen
  \bibfield  {author} {\bibinfo {author} {\bibfnamefont {M.}~\bibnamefont {Nadeem}}, \bibinfo {author} {\bibfnamefont {M.~S.}\ \bibnamefont {Fuhrer}},\ and\ \bibinfo {author} {\bibfnamefont {X.}~\bibnamefont {Wang}},\ }\bibfield  {title} {\bibinfo {title} {The superconducting diode effect},\ }\href@noop {} {\bibfield  {journal} {\bibinfo  {journal} {Nature Reviews Physics}\ }\textbf {\bibinfo {volume} {5}},\ \bibinfo {pages} {558} (\bibinfo {year} {2023}{\natexlab{a}})}\BibitemShut {NoStop}%
\bibitem [{\citenamefont {Nadeem}\ \emph {et~al.}(2023{\natexlab{b}})\citenamefont {Nadeem}, \citenamefont {Fuhrer},\ and\ \citenamefont {Wang}}]{nadeem23}%
  \BibitemOpen
  \bibfield  {author} {\bibinfo {author} {\bibfnamefont {M.}~\bibnamefont {Nadeem}}, \bibinfo {author} {\bibfnamefont {M.~S.}\ \bibnamefont {Fuhrer}},\ and\ \bibinfo {author} {\bibfnamefont {X.}~\bibnamefont {Wang}},\ }\bibfield  {title} {\bibinfo {title} {Superconducting diode effect--fundamental concepts, material aspects, and device prospects},\ }\href@noop {} {\bibfield  {journal} {\bibinfo  {journal} {arXiv preprint arXiv:2301.13564}\ } (\bibinfo {year} {2023}{\natexlab{b}})}\BibitemShut {NoStop}%
\end{thebibliography}%

\begin{widetext}
\begin{center}
\LARGE \textbf{Supplementary Information}
\end{center}

\begin{center}
\Large \textbf{Electronic dispersion and the role of entropy}
\end{center}
Band dispersion of zero-entropy VCl$_3$ monolayer, 1 x 1 unit cell and 2 x 2 supercell, and entropic VCl$_3$ monolayer with long-range order (LRO) along the zigzag chains [ZZ-ttt] is shown in figure \textbf{S1}. Unlike half-metallic character in zero-entropy monolayers, figure \textbf{S1}(a-d), ZZ-ttt entropic configuration exhibits spin gapless semiconducting (SGS) behavior, where SOI opens a nontrivial energy gap leading to a fully gapped quantum anomalous Hall (QAH) phase, figure \textbf{S1}(e,f). Figure \textbf{S2} shows band dispersion for other three LRO-entropic configurations, namely SL-t$^\prime$t$^\prime$t$^\prime$], Mix-t$^\prime$t$^\prime$t, and Mix-t$^\prime$tt. 

In the LRO-entropic chiral structure [SL-t$^\prime$t$^\prime$t$^\prime$], TM-atoms are substituted on the sublattice sites such that the triangular sublattice A is formed by V atoms and the triangular sublattice B is formed by M$^\prime$ (Ti, Cr, Fe, Co) atoms. In this configuration, an increase in entropy modifies nearest neighbor hopping between V atoms and the dopant M$^\prime$ atoms such that a gap is induced in the low-energy bands. As a consequence, the LRO-entropic chiral structure displays a gapped dispersion featuring a ferromagnetic insulating phase, as shown in figure \textbf{S2}(a). Consistent with previous analysis for the ZZ-ttt configuration, entropy engineering drives band flattening along with a blue shift at K/K$^\prime$ points while a red shift across M-point. However, the bulk states across the $\Gamma$-point remain mostly unaffected, indicating a redistribution of various orbitals' contributions to the low-energy bands and the associated crystal fields. Furthermore, unlike the ZZ-ttt configuration that displays an intertwining between localization and de-localization of d-orbitals, localization dominates over de-localization in the SL-t$^\prime$t$^\prime$t$^\prime$ configuration. On the other hand, in the Mix-t$^\prime$t$^\prime$t LRO entropic configuration, an interplay between localization and de-localization leads to the band dispersion that exhibits a nodal-line semi-metallic character between nearly flat spin-up and spin-down bands, as shown in figure \textbf{S2}(b). Interestingly, a small change in this configuration, induced by exchanging a sublattice site of one of the V atoms with a Cr atom, drastically changes the band dispersion, as shown in figure \textbf{S2}(c). This shows that entropy engineering is an effective mechanism for controlling the bandstructure.

Detailed orbital resolved band dispersion of zero-entropy VCl$_3$ monolayer (1 x 1 unit cell) and the LRO-entropic VCl$_3$ monolayer ZZ-ttt (2 x 2 unit cell) are shown in figure \textbf{S3} and \textbf{S4}, respectively. In the zero-entropy VCl$_3$ monolayer, the low-energy conduction band of the Dirac dispersion is predominantly occupied by the d$_{x^2-y^2}$ orbital, along with a contribution from the d$_{xy}$ and d$_{xy}$ orbitals along the M-K-$\Gamma$ and K-M-$\Gamma$ line, respectively. On the other hand, low-energy valence band of the Dirac dispersion is predominantly occupied by the d$_{z^2}$ orbital, along with a small contribution from in-plane 3d orbitals (d$_{x^2-y^2}$ and d$_{xy}$) around the Dirac point. However, in the LRO-entropic ZZ-ttt configuration, the low-energy Dirac bands are completely occupied by the in-plane 3d orbitals (d$_{x^2-y^2}$ and d$_{xy}$) while the 3d orbitals with out-of-plane component (d$_{z^2}$, d$_{yz}$, d$_{zx}$) do not contribute to the states around the Fermi level in the spin-up channel. Orbital contributions from doped M$^\prime$ atoms are also depicted in figure \textbf{S4}.

\begin{center}
\Large \textbf{Magnetic moments and local spin textures}
\end{center}
As shown in tables \ref{ZE} and \ref{LRO}, local magnetic moments and local spin textures strongly depend on the level of entropy and the nearest-neighbor environment. In addition, the entropy-induced transformation of the ferromagnetic ground state, as well as the modification of electronic structure, can also be understood based on the differences between the number of valence electrons contributed by the M$^\prime$ atoms and those from the V atoms. For instance, similar to the ferromagnetic ground state of the VCl$_3$ monolayer with a total magnetic moment of 4 $\mu_B$ (16 $\mu_B$ for a 2 x 2 supercell), the ground states of the LRO-entropic monolayers ZZ-ttt and SL-t$^\prime$t$^\prime$t$^\prime$ remain ferromagnetic, but the total magnetic moment reduces to 10.972 $\mu_B$ and 12.143 $\mu_B$, respectively. In addition, local magnetic moments of individual TM cations and local spin textures strongly depend on the level of entropy and the type of LRO. 

In the LRO-entropic monolayer ZZ-ttt, similar to the high-spin state of vanadium V$^{3+}$ [$3d^2$], the magnetic moments of Ti ($m_B\approx+0.413$ $\mu_B$) and Cr ($m_B\approx+2.738$ $\mu_B$) are found to be consistent with their respective values in the +3 oxidation state, i.e., Ti$^{3+}$ [$3d^1$] and Cr$^{3+}$ [$3d^3$]. However, the magnetic moments of Co ($m_B\approx+0.582$ $\mu_B$) suggest that it is found in the low-spin state, Co$^{2+}$ [$3d^7$]. On the other hand, the Fe atom contributes $m_B\approx-0.743$ $\mu_B$ to the total magnetic of the ZZ-ttt configuration, which is oppositely aligned to the magnetic moments of other TM cations. It suggests that Fe is found to be in the low-spin state Fe$^{3+}$ [$3d^5$] and the Cl$\downarrow$-Fe$\downarrow$-Cl$\downarrow$ bond favors ferromagnetic interactions. This behavior of Fe in the LRO-entropic monolayer ZZ-ttt is completely different from that of Fe cations in Fe-doped low-entropy VCl$_3$ \cite{ouettar23}, where Fe$^{3+}$ [$3d^5$] is found to exist in a high-spin state ($m_B\approx+4.106$ $\mu_B$), enhancing the total magnetic moment of the VCl$_3$ 2 x 2 monolayer from 16 $\mu_B$ to 19 $\mu_B$, and the ferromagnetic nature of Cl$\uparrow$-Fe$\uparrow$-Cl$\uparrow$ bond is indebted to the positive value of the local magnetic moment of the Cl atoms. The local spin texture of Fe and the deviation from a high-spin state in the low-entropy case to a low-spin state in the high-entropy case could be a consequence of the level of entropy and symmetry-breaking effects.  

In the SL-t$^\prime$t$^\prime$t$^\prime$ configuration, low/high spin states of the TM cations remain the same as in the ZZ-ttt configuration. However, unlike the ZZ-ttt configuration, the magnetic moments of all the TM cations remain positive, and thus an antiferromagnetic interaction is favored by all the Cl$\downarrow$-TM$\uparrow$-Cl$\downarrow$ bonds, as indicated by the negative local magnetic moments of the Cl atoms. That is, the local spin textures are similar to that in the zero-entropy case. However, the magnetic moments of V atoms are further reduced from the zero-entropy case.

Local magnetic moments and local spin textures become more interesting in the LRO-entropic Mix-t$^\prime$t$^\prime$t and Mix-t$^\prime$tt configurations. First, the magnetic moments of two of the V atoms are significantly reduced to $+0.393$ $\mu_B$ and $+0.253$ $\mu_B$ in the Mix-t$^\prime$t$^\prime$t configuration. Second, like the ZZZ-ttt and SL-t$^\prime$t$^\prime$t$^\prime$ configurations, Fe and Cr favor a low-spin state and a high-spin state, respectively. However, the local magnetic moments of both Fe and Cr are negative. Third, unlike the ZZZ-ttt and SL-t$^\prime$t$^\prime$t$^\prime$ configurations, the magnetic moments of Co ($m_B\approx+2.455$ $\mu_B$) suggest that it is found in a high-spin state, Co$^{2+}$ [$3d^7$]. 

In the LRO-entropic Mix-t$^\prime$tt, on the other hand, the magnetic moment of only one of the V atoms is reduced to $+0.421$ $\mu_B$ while the other three shows $m_B\approx+1.9$ $\mu_B$. Interestingly, the reduced magnetic moment of that V atom appears to be negative, highly contrasting from other LRO-entropic configurations. In addition, while the magnetic moment of the low-spin Fe state remains negative, the magnetic moment of the high-spin Cr state becomes positive. Furthermore, unlike Mix-t$^\prime$t$^\prime$t but like ZZZ-ttt and SL-t$^\prime$t$^\prime$t$^\prime$ configurations, the magnetic moments of Co ($m_B\approx+0.597$ $\mu_B$) suggest that it is found in the low-spin state, Co$^{2+}$ [$3d^7$]. The magnetic ground state of Mix-t$^\prime$t$^\prime$t and Mix-t$^\prime$tt configurations stabilizes with a reduced total magnetic moment of $m_B=+3.286$ $\mu_B$ and $m_B=+8.747$ $\mu_B$, respectively. In both of these mix entropic configurations, positive magnetic moments for some of the Cl atoms while negative magnetic moments for the other Cl atoms suggest an intermingling of ferromagnetic and antiferromagnetic Cl$\uparrow\downarrow$-TM$\uparrow\downarrow$-Cl$\uparrow\downarrow$ bonds.

A momentous change in the local magnetic moments and spin textures in the LRO-entropic Mix-t$^\prime$t$^\prime$t and Mix-t$^\prime$tt configurations show that the electronic and magnetic properties are significantly altered by the relocation of a single Cr atom, indicating the fragile impact of entropy engineering on electronic and magnetic properties.

\begin{table*}
\centering
\begin{tabular}{cccccccccccc}
Sample&M$_{V_1}$&M$_{V_2}$&M$_{V_3}$&M$_{V_4}$&M$_{V_5}$&M$_{V_6}$&M$_{V_7}$& M$_{V_8}$&M$_{Cl}$&M$_{total}$\\
\hline
VCl$_3$(1x1) &$1.935\uparrow$&$1.935\uparrow$&-&-&-&-&-&-&$\sim0.03\downarrow$&4$\uparrow$\\
VCl$_3$(2x2)&$1.916\uparrow$&$1.916\uparrow$&$1.916\uparrow$&$1.916\uparrow$&$1.916\uparrow$&$1.916\uparrow$&$1.916\uparrow$&$1.916\uparrow$&$\sim0.03\downarrow$&16$\uparrow$\\
\end{tabular}
\caption{Magnetic moments and spin orientations of V-atoms in zero-entropy VCl$_3$ monolayer. All values are given in $\mu_B$.}
\label{ZE}
\end{table*}

\begin{table*}
\centering
\begin{tabular}{cccccccccccc}
Sample&M$_{V_1}$&M$_{V_2}$&M$_{V_3}$&M$_{V_4}$&M$_{Ti}$&M$_{Fe}$&M$_{Cr}$& M$_{Co}$&M$_{Cl}$&M$_{total}$\\
\hline
ZZ-ttt&
$1.918\uparrow$&$1.876\uparrow$&$1.822\uparrow$&$1.783\uparrow$& $0.413\uparrow$ &\cellcolor[HTML]{D3D3D3}$0.743\downarrow$ & $2.738\uparrow$&$0.582\uparrow$&(0.001-0.034)$\downarrow$&$10.972\uparrow$\\
SL-t$^\prime$t$^\prime$t$^\prime$& $1.863\uparrow$&1.849$\uparrow$&1.779$\uparrow$&1.697$\uparrow$& $0.485\uparrow$ & $0.533\uparrow$ & $2.748\uparrow$&$0.552\uparrow$&(0.001-0.036)$\downarrow$&$12.143\uparrow$\\
\cellcolor[HTML]{D3D3D3}Mix-t$^\prime$t$^\prime$t& 
1.827$\uparrow$&1.762$\uparrow$&\cellcolor[HTML]{D3D3D3}0.393$\uparrow$&\cellcolor[HTML]{D3D3D3}0.253$\uparrow$& $0.166\uparrow$&\cellcolor[HTML]{D3D3D3}$0.963\downarrow$& \cellcolor[HTML]{D3D3D3}$2.713\downarrow$& \cellcolor[HTML]{D3D3D3}2.455$\uparrow$&\cellcolor[HTML]{D3D3D3}(0.001-0.035)$\downarrow\uparrow$&$03.286\uparrow$\\
\cellcolor[HTML]{D3D3D3}Mix-t$^\prime$tt& 1.948$\uparrow$&1.946$\uparrow$&1.902$\uparrow$&\cellcolor[HTML]{D3D3D3}0.421$\downarrow$& $0.402\uparrow$&\cellcolor[HTML]{D3D3D3}$0.849\downarrow$&$2.777\uparrow$&$0.597\uparrow$& \cellcolor[HTML]{D3D3D3}(0.001-0.033)$\downarrow\uparrow$& $08.747\uparrow$\\
\end{tabular}
\caption{Magnetic moments and spin orientations of TM cations in LRO-entropic TiV$_4$CrFeCoCl$_{24}$ monolayers. Shaded cells represent an entropy-driven transition in spin up/down polarization and/or low/high spin states of TM atoms and the appearance of Cl atoms in both spin-up and spin-down states. All values are given in $\mu_B$.}
\label{LRO}
\end{table*}

\begin{figure*}[ht]
\centering
\includegraphics[width=0.9\linewidth]{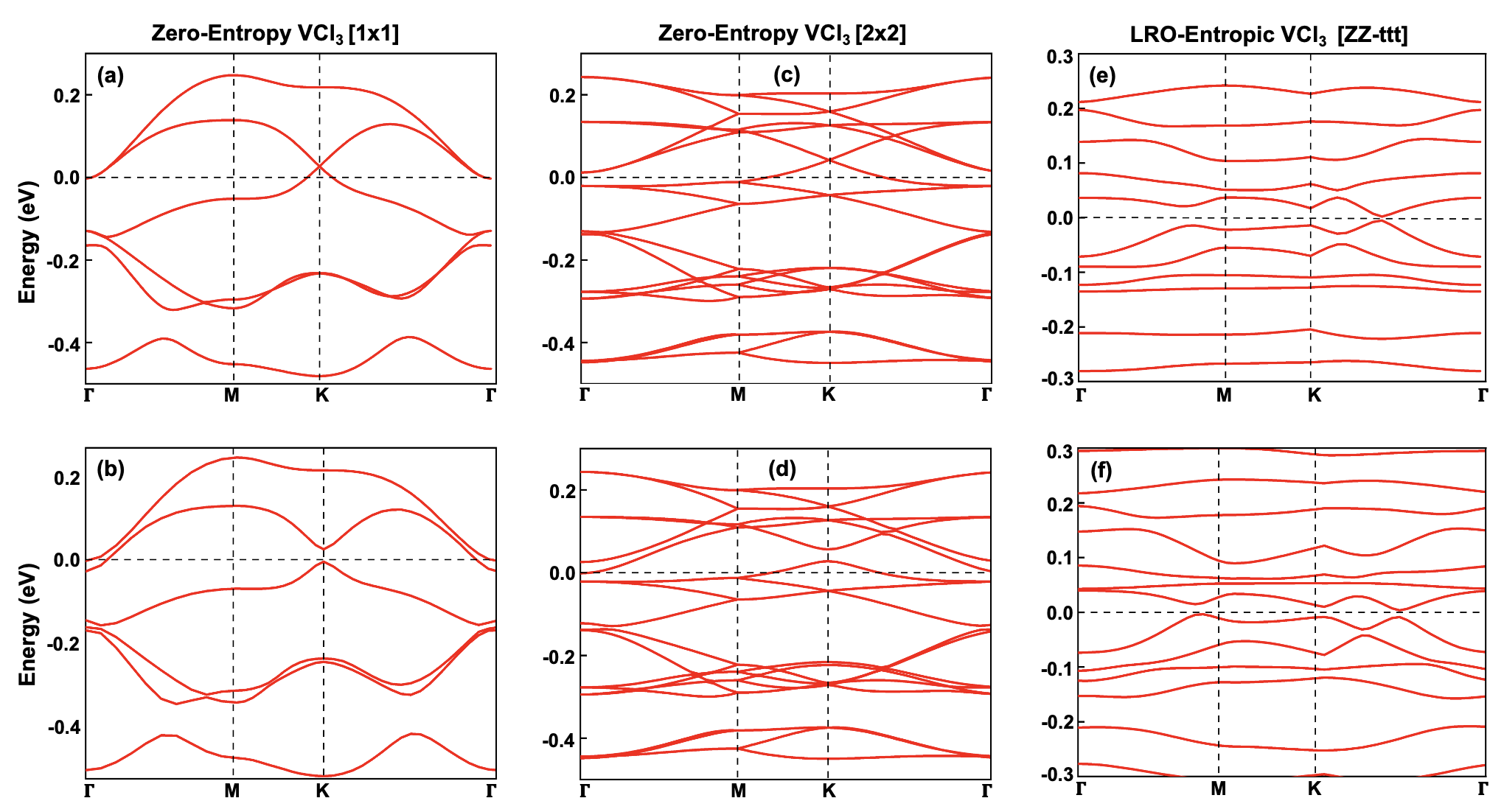}
\caption{\textbf{[S1:] Electronic dispersion of vanadium trichloride monolayer.} \textbf{(a,b)} Band dispersion of zero-entropy VCl$_3$ monolayer (1 x 1 unit cell) without SOI (a) and with SOI (b). \textbf{(c,d)} Band dispersion of zero-entropy VCl$_3$ monolayer (2 x 2 supercell) without SOI (c) and with SOI (d). \textbf{(e,f)} Band dispersion of entropic TiV$_4$CrFeCoCl$_{24}$ monolayer (2 x 2 supercell), with a long-range order along zigzag chains, without SOI (e) and with SOI (f).}
\label{2DBS}
\end{figure*}

\begin{figure*}[ht]
\centering
\includegraphics[width=0.9\linewidth]{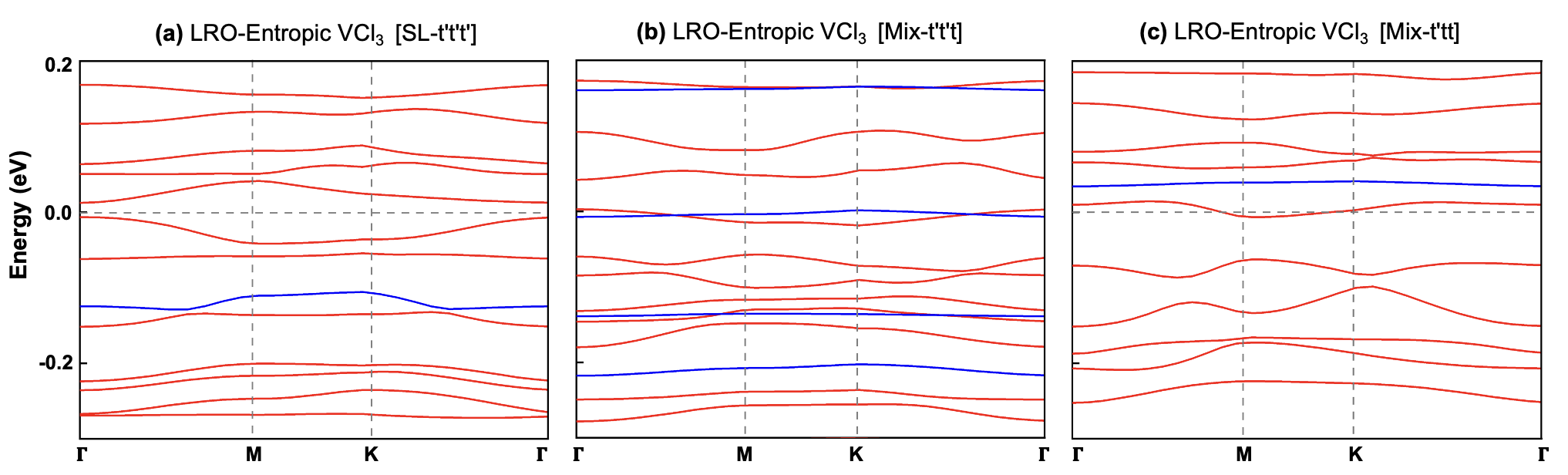}
\caption{\textbf{[S2:] Nearest neighbor effect on the electronic dispersion of LRO entropic VCl$_3$ monolayer.} Band dispersion of LRO entropic VCl$_3$ monolayer [SL-t$^\prime$t$^\prime$t$^\prime$] (a) [Mix-t$^\prime$t$^\prime$t] (b), and [Mix-t$^\prime$tt. Here red and blue bands represent spin-up and spin-down sectors, respectively.]}
\label{LRO-BS}
\end{figure*}

\begin{figure*}[ht]
\centering
\includegraphics[width=0.9\linewidth]{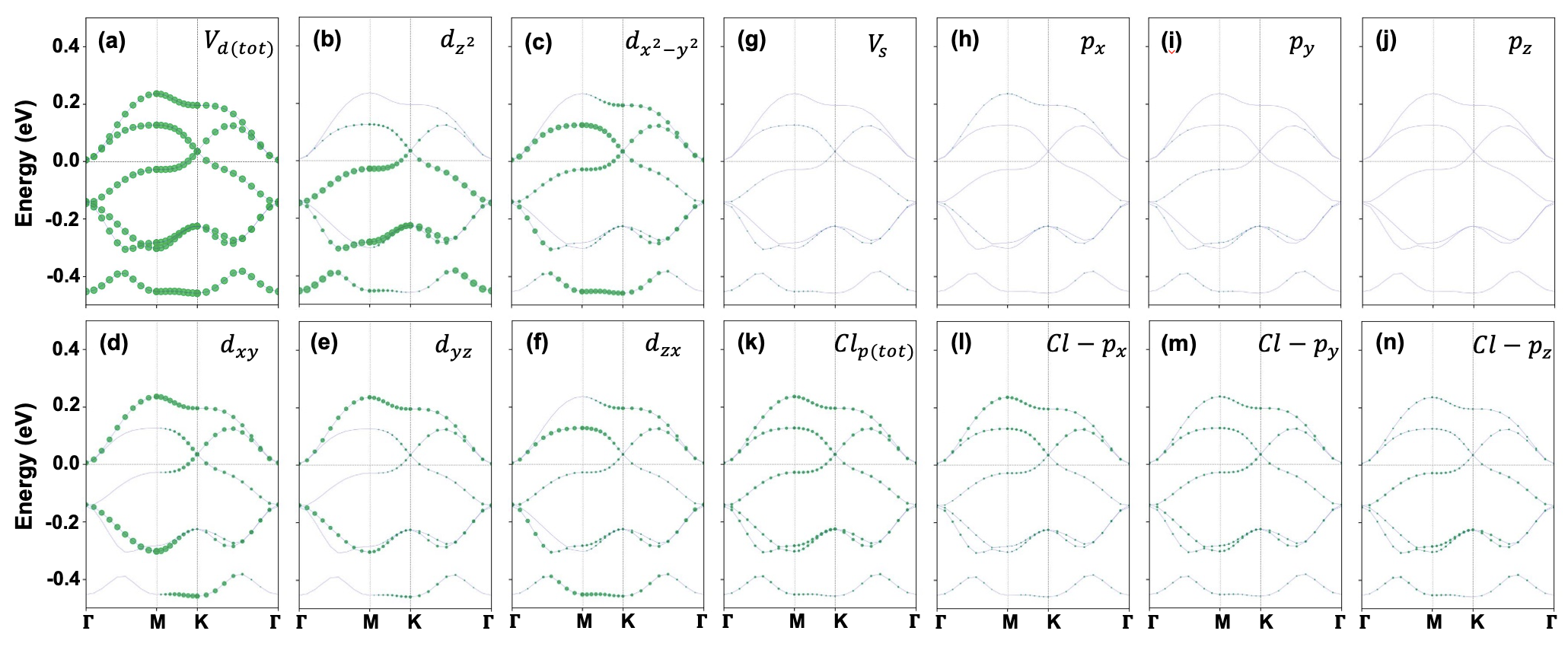}
\caption{\textbf{[S3:] Orbital resolved band dispersion of VCl$_3$ monolayer.} \textbf{(a-f)} Orbital contribution from d-orbitals of vanadium (a), $e_g$ orbitals (d$_{z^2}$, d$_{x^2-y^2}$) (b,c) and $t_{2g}$ orbitals (d$_{xy}$, d$_{yz}$, d$_{zx}$) (d,e,f). \textbf{(g-j)} Orbital contribution from s-orbitals (g) and p-orbitals (h,i,j) of vanadium. \textbf{(k-n)} Orbital contribution from p-orbitals of Cl. The size of the bands represents the orbital weight.}
\label{ORBD-ZE}
\end{figure*}

\begin{figure*}[ht]
\centering
\includegraphics[width=0.9\linewidth]{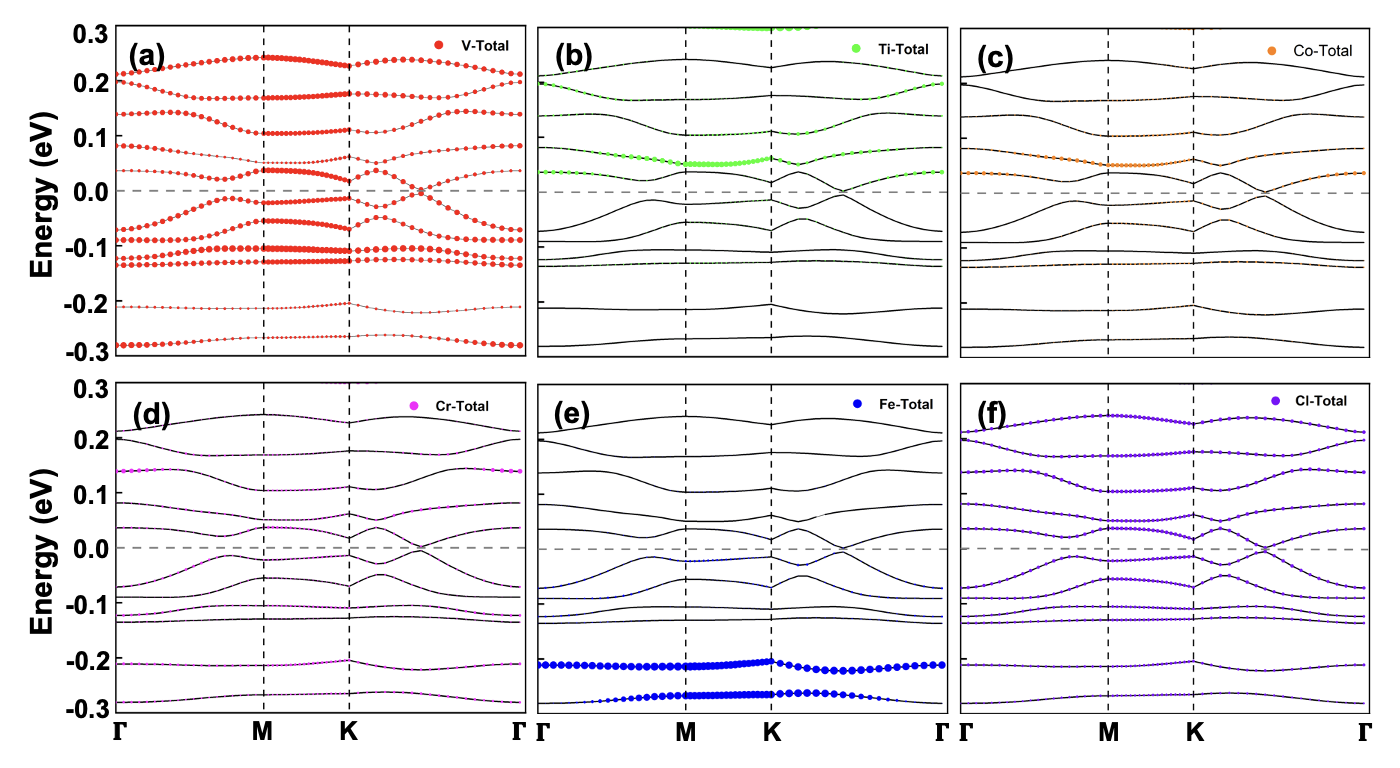}
\caption{\textbf{[S4:] Orbital resolved band dispersion of entropic VCl$_3$ monolayer.} \textbf{(a-f)} Orbital-resolved band structures in TiV$_4$CrFeCoCl$_24$ monolayer with the corresponding contribution from V (a), Ti (b), Co (c), Cr (d), Fe (e), and Cl (f) atoms, respectively. The size of the bands represents the orbital weight.}
\label{ORBD}
\end{figure*}

\end{widetext}
\end{document}